\documentclass[aps,prb,letterpaper,superscriptaddress,10pt,twocolumn,floatfix]{revtex4-1}
\usepackage[colorlinks=true,allcolors=blue]{hyperref}
\usepackage{amssymb}
\usepackage{amsmath}
\usepackage{float}
\usepackage{graphicx}
\usepackage[caption=false]{subfig}
\usepackage{amsfonts}
\usepackage{xcolor}
\usepackage{epsfig}
\usepackage{color}
\usepackage{bm}
\usepackage{tabularx}
\usepackage{multirow}
\usepackage{mathtools}
\usepackage{xfrac}
\usepackage{blkarray}
\usepackage{bbold}
\usepackage[mathscr]{euscript}
\usepackage{autobreak}
\usepackage{comment}
\usepackage{makecell}
\usepackage{soul, color, xcolor}
\soulregister{\cite}7 
\soulregister{\citep}7 
\soulregister{\citet}7
\soulregister{\ref}7
\soulregister{\onlinecite}7
\usepackage{float}
\begin{document}
\title{Carbon cluster emitters in silicon carbide}

\author{Pei Li}
\affiliation{Beijing Computational Science Research Center, Beijing 100193, China}
\affiliation{Wigner Research Centre for Physics, P.O.\ Box 49, H-1525 Budapest, Hungary}

\author{P\'eter Udvarhelyi}
\affiliation{Wigner Research Centre for Physics, P.O.\ Box 49, H-1525 Budapest, Hungary}
\affiliation{Department of Atomic Physics, Institute of Physics, Budapest University of Technology and Economics, M\H{u}egyetem rakpart 3., H-1111 Budapest, Hungary}

\author{Song Li}
\affiliation{Wigner Research Centre for Physics, P.O.\ Box 49, H-1525 Budapest, Hungary}

\author{Bing Huang}
\affiliation{Beijing Computational Science Research Center, Beijing 100193, China}
\affiliation{Department of Physics, Beijing Normal University, Beijing, 100875, China}

\author{Adam Gali}
\affiliation{Wigner Research Centre for Physics, P.O.\ Box 49, H-1525 Budapest, Hungary}
\affiliation{Department of Atomic Physics, Institute of Physics, Budapest University of Technology and Economics, M\H{u}egyetem rakpart 3., H-1111 Budapest, Hungary}

\date{\today}
%====================================================

%-ABSTRACT-------------------------------------------
\begin{abstract}
%Silicon carbide in its 4H polytype (4H-SiC) is a promising wide band gap semiconductor for highly-demanding electronic devices, thanks to its high breakdown electrical field, high carrier saturation speed, excellent thermal conductivity, and other favorable properties. 
Defect qubits in 4H-SiC are outstanding candidates for numerous applications in the rapidly emerging field of quantum technology. Carbon clusters can act as emission sources that may appear after thermal oxidation of 4H-SiC or during irradiation which kicks out carbon atoms from their sites. These fluorescent carbon clusters could interfere with the already established vacancy-related qubits that generated with irradiation techniques. In this study, we systematically investigate the electronic structure, formation energy, dissociation energy, vibrational properties, and the full fluorescence spectrum of carbon clusters involving up to four carbon atoms in 4H-SiC by means of density functional theory calculations. All the possible local configurations for these carbon clusters are carefully evaluated. We find the electronic and vibronic properties of the carbon clusters depend strongly on the local configuration of the 4H-SiC lattice. By comparing the calculated and previously observed fluorescence spectra in 4H-SiC, we identify several carbon clusters as stable visible emitters in 4H-SiC. The paired carbon interstitial defects are identified as the source of the 463-nm triplet and the 456.6-nm emitters. The 471.8-nm emitter in 4H-SiC is associated with tri-carbon antisite clusters. Our findings provide plausible explanation for the origin of visible emission lines in 4H-SiC and propose the possible configurations of carbon clusters which are helpful for the quantum information processing application through qubits in 4H-SiC.
\end{abstract}

\maketitle
%====================================================

%-INTRODUCTION---------------------------------------
\section{Introduction}
\label{sec:introduction}
Fluorescent paramagnetic point defects in solids have attracted a great deal of attention as they can act as single-photon sources and quantum bits~\citep{wolfowicz2021quantum} that are the base of realizing quantum information processing applications. 

Silicon carbide (SiC), which is an established material for semiconductor industry, has been considered as an attractive material for quantum information processing. In particular, the 4H polytype of SiC (4H-SiC) with the fundamental band gap of 3.23~eV at cryogenic temperatures acts as a favorable host for color centers, which is the most advanced polytype among the SiC polytypes from technology point of view~\citep{SiC-review-paper-Son-Awschalom, SiC-csore-review-paper-for-defects}. In 4H-SiC, there are two types of Si-C bilayer arrangements called as hexagonal ($h$) and quasi-cubic ($k$) sites, thus a single type of substitutional or vacancy defect has two configurations, $h$ and $k$ configurations, whereas pairs of substitutional or vacancy defects have four different configurations. For interstitial defects, one may also label the configurations according to the position of the interstitial atom in the corresponding Si-C bilayer inside 4H-SiC. 

Several defects in 4H-SiC are reported to act as a single photon source operating up to room temperature such as silicon-vacancy ($\text{V}_{\text{Si}}$) with the characteristic zero-phonon-lines (ZPLs) at 1.438 and 1.352~eV~\citep{ivady2017identification,wagner2000electronic} and divacancy ($\text{V}_{\text{Si}}$-$\text{V}_{\text{C}}$) with the ZPLs at 1.095, 1.096, 1.119 and 1.150 eV~\citep{falk2013polytype,csore2022fluorescence}. They exhibit several desirable magneto-optical properties which makes it possible to coherently manipulate single spins~\citep{koehl2011room,widmann2015coherent,christle2015isolated}. 

Beside these solid-state defect qubits, other photoluminescence (PL) centers, e.g. the $D_{\text{I}}$~\citep{egilsson1999properties}, $D_{\text{II}}$~\citep{sullivan2007study, patrick1973localized}, and the 463-nm triplet~\citep{steeds2008creation,steeds2008identification}, were detected in electron irradiated and implanted 4H-SiC samples which were associated with native defects~\citep{gali2003correlation, mattausch2004structure}. By the calculated single-electron level position and (quasi) local vibration modes, the involvement of silicon antisite is suggested for the $D_{\text{I}}$ center~\citep{gali2003correlation, eberlein2006density} which exhibits ZPL emission at 2.901~eV in 4H-SiC~\citep{egilsson1999properties}. For $D_{\text{II}}$ center with ZPL emission at 3.205~eV, several carbon cluster models have been suggested~\citep{gali2003aggregation, mattausch2004structure, jiang2012carbon}, based on the localized vibrational mode (LVM) frequencies~\citep{sullivan2007study}. The properties of the 463-nm triplet emitters in 4H-SiC~\citep{steeds2008creation}, where the 463-nm refers to the position of the ZPL wavelength of about 2.68~eV, show great similarities to the so-called \textit{P-T} centers observed in electron irradiated 6H-SiC, where the \textit{P-T} centers have ZPL energy at around 2.45~eV~\citep{evans2002identification}. We note that 6H-SiC has about 0.23~eV smaller band gap than that of 4H-SiC, and it has two quasi-cubic sites which makes the number of possible defect configurations larger than those in 4H-SiC. Nevertheless, the formation conditions of these centers suggest that the 463-nm triplet color center in 4H-SiC has the same origin as the \textit{P-T} centers in 6H-SiC, where the latter PL centers were associated with carbon split interstitials or di-carbon antisite defects by comparing the calculated and observed LVMs in the phonon sideband~\citep{mattausch2004structure, evans2002identification, gali2003aggregation, mattausch2006thermally}. In electron-irradiated 6H-SiC, the \textit{U} (2.360~eV) and \textit{Z} (2.417~eV) centers were also observed~\citep{evans2002identification} which were associated with tri-carbon antisite and di-carbon interstitial cluster, respectively~\citep{gali2003aggregation, mattausch2006thermally}, again based on the calculated and observed LVMs of the given carbon clusters.    

The idea behind these carbon-related models was based on the low-barrier migration energy of the carbon interstitials~\citep{bockstedte2003ab,rauls2001interstitial,PhysRevB.69.235202} where the migrating carbon interstitials may join to form carbon clusters. These clusters can act as sinks for carbon self-interstitials at lower temperatures (aggregation), and they can re-emit carbon self-interstitials at higher temperatures (dissociation). In these theoretical studies~\citep{gali2003aggregation, mattausch2004structure, mattausch2006thermally, jiang2012carbon}, only the high energy LVMs were calculated and compared to the features in the PL spectra associated with the LVMs of the PL centers. However, it is known that the intensity of the LVMs in the PL spectra depends on the change of the geometry reconfiguration upon optical excitation, thus the association based on solely the calculated LVMs is ambiguous. Furthermore, the possible variations and stability of the carbon clusters in different local configurations of 4H-SiC crystal, i.e., alternating $h$ and $k$ configurations, were often ignored in previous theoretical studies.  

In this work, we systematically investigate clusters of carbon interstitials and antisites in 4H-SiC. To this end, we calculate the defect level (DL), formation energy, dissociation energy, vibrational properties, fluorescence spectrum with ZPL energy and phonon sideband where we focus our study on the near-infrared and visible carbon-related emitters. Therefore, identification of $D_\text{I}$ and $D_\text{II}$ fluorescent centers is out of the scope of the present study. Our paper is organized as follows. We describe the computational methodology in Sec.~\ref{sec:method}, including the methodology of the \textit{ab initio} density functional theory calculations, briefly description of formation energy, dissociation energy, and Huang-Rhys theory. In Sec.~\ref{sec:result}, we present the results for interstitial and antisite clusters in detail and illustrate the rich variety of possible defect structures. In Sec.~\ref{sec:discussion}, we discuss the correlation between our calculations and the previously reported PL centers in 4H-SiC, 6H-SiC and $\text{4H-SiC}/\text{SiO}_2$ interface.   
%====================================================

%-Method---------------------------------------------
\section{Methods}
\label{sec:method}
%-ini---vasp------
\subsection{Computational methodology}
All the first-principles calculations are performed using density functional theory (DFT) within the projector augmented wave potential plane-wave method, as implemented in the Vienna ab-initio simulation package (VASP 5.4.1)~\citep{PhysRevB.54.11169} with the projector augmented wave method~\citep{PhysRevB.50.17953}. The electron wave functions are expanded in plane-wave basis set limited by a cutoff of 420~eV. The fully relaxed geometries were obtained by minimizing the quantum mechanical forces between the ions falling below the threshold of 0.01~eV/\AA\ and the self-consistent calculations are converged to $10^{-5}$~eV.

The screened hybrid density functional of Heyd, Scuseria, and Ernzerhof (HSE)~\citep{heyd2003hybrid} is employed to calculate the electronic structure. In this approach, we could mix part of nonlocal Hartree-Fock exchange to the generalized gradient approximation of Perdew, Burke, and Ernzerhof (PBE)~\citep{PhysRevLett.77.3865} with the default fraction ($\alpha$ = 0.25) and the inverse screening length at 0.2~\AA$^{-1}$. The calculated band gap at 3.17~eV reproduces well the experimental value at 3.23~eV in Refs.~\onlinecite{SiC-review-paper-Son-Awschalom, SiC-csore-review-paper-for-defects}. We embedded the carbon clusters in a 576-atom 4H-SiC supercell which is sufficient to minimize the periodic defect-defect interaction. The single $\Gamma$-point scheme is convergent for the k-point sampling of the Brillouin zone. The excited states were calculated by $\Delta$SCF method~\citep{PhysRevLett.103.186404}. The migration barrier is determined by using the climbing image nudged elastic band (CI-NEB) methods~\citep{mills1995reversible,henkelman2000climbing,henkelman2000improved}.  

For relevant paramagnetic defects, we determined the hyperfine interaction between the electron spin and the $^{13}$C and $^{29}$Si $I=1/2$ nuclear spins which consists of the Fermi-contact term with core-polarization correction and dipole-dipole term as implemented in VASP~\citep{szasz2013hyperfine}. The electron spin-electron spin dipole-dipole interaction was calculated for the defects with $S=1$ electron spin as implemented by Martijn Marsman in VASP within the spinpolarized Kohn-Sham DFT~\cite{Bodrog2013, Ivady2014, Ivadyreview}.  In this case, the axial $D$ parameter and the orthorhombic $E$ parameter are provided. 

For the vibrational modes, we calculated the corresponding dynamical matrix containing the second order derivatives of the total energy by means of the PBE~\citep{PhysRevLett.77.3865} functional where all the atoms in the supercell were enabled to vibrate. In this case, we apply strict threshold parameters for the convergence of the electronic structure ($10^{-6}$~eV) and atomic forces ($10^{-3}$~eV/\AA ) in the geometry optimization procedure. We found that the calculated local vibration modes in a 256-atom and 576-atom supercells agree within 3~meV, and we conclude that the 576-atom supercell size is sufficiently large to obtain accurate local vibration modes above the top Raman-frequency of 4H-SiC. These vibration modes are applied together with the HSE ground state and excited state geometries to simulate the PL spectrum of the given defects within Huang-Rhys theory as explained below. This strategy worked well for deep defects in diamond (e.g., Refs.~\onlinecite{ThieringNV2017, ThieringG4V2018}).

%-end---vasp-----------------

%-ini---Formation energy------
\subsection{Formation energy}
The formation energies $E_f$ of defects are calculated to determine the charge stability with the following equation~\citep{PhysRevLett.67.2339, RevModPhys.86.253},
\begin{equation}
\label{equation1}
\begin{split}
E^q_f = &E^q_d - E_\text{host}
+ \sum n_i(\Delta \mu_i + E_i)\\
&+ q\left(E_\text{V} + E_\text{Fermi}\right) + E_\text{corr}\left(q\right)\text{,}
\end{split}
\end{equation}
\noindent where $E^q_d$ is the total energy calculated with a supercell containing a defect in a charge state $q$. $E_\text{host}$ is the total energy for the defect-free supercell, whereas $n_i$ is the number of atoms removed from or added to the supercell. $\Delta \mu_i$ is the chemical potential of constituent $i$ referenced to bulk silicon and diamond with energy $E_i$. Generally, SiC growth can be performed under different experimental conditions, e.g., from Si-rich to C-rich conditions. By varying the chemical potentials in the calculation, different experimental scenarios can be explored. In this paper, we only consider the carbon-rich condition for the sake of simplicity. $E_\text{V}$ and $E_\text{Fermi}$ are the energy of the valence band maximum (VBM) and Fermi-level referenced to the VBM of the host, respectively. The $E_\text{corr}\left(q\right)$ is the correction term for the charged system due to the existence of electrostatic interactions of the periodic images of the defect. We used the correction method of Freysoldt, Neugebauer and Van de Walle (FNV)~\cite{FNV} for the total energy of the defective systems. 

We note that we also applied FNV corrections for such excited states of the neutral defects where the excited state has a pseudo-donor nature, i.e., the system can be considered as a positively charged core and a loosely bound electron orbiting in the effective mass state. The background for the need of the size correction following the charge correction of a charged supercell is explained in Ref.~\onlinecite{Gali2023} that has been successfully applied to neutral silicon-vacancy in diamond~\cite{Zhang2020} and the carbon-oxygen split interstitial cluster in silicon~\cite{Udvarhelyi2022}.

The charge transition level $\epsilon^{q/q^{\prime}}_d$ can be derived from Eq.(\ref{equation1}) as follows
\begin{equation}
\label{equation2}
\begin{split}
\epsilon^{q/q^{\prime}}_d = (E^q_f - E^{q^{\prime}}_f)/(q^{\prime} - q)\text{,}
\end{split}
\end{equation}
\noindent where $E^q_f$ and $E^{q^{\prime}}_f$ are the formation energy of the defect in charge state $q$ and $q^{\prime}$, respectively.
%-end---Formation energy------

%-ini---dissociation energy------
\subsection{Dissociation energy}
The formation energy can be used as the standard to judge the abundance of a defect under equilibrium conditions. However, the defects formation during irradiation is not an equilibrium process and the calculated formation energy is not a decisive factor. To provide a more comprehensive assessment of defect stability, the dissociation energy of the carbon cluster is calculated.

The dissociation energy $E_D$, which is the energy required to remove a single carbon atom from the carbon cluster could be defined with the following equation,
\begin{equation}
\label{equation3}
\begin{split}
E_D = &E_\text{tot}(\text{C}_{n-1}) + E_\text{tot}(\text{C}_\text{sp}) - E_\text{tot}(\text{C}_{n})\text{,}
\end{split}
\end{equation}
\noindent where $E_\text{tot}(\text{C}_{n-1})$ is the total energy of the carbon cluster that results from the removal of a carbon atom. $E_\text{tot}(\text{C}_\text{sp})$ is the total energy of the carbon split-interstitial at the appropriate site in 4H-SiC, which is the most stable configuration of the isolated carbon defect~\citep{mattausch2004carbon}. $E_\text{tot}(\text{C}_{n})$ is the total energy of the initial carbon cluster prior to the removal of the carbon atom.
%-end---Dissociation energy------

%-ini---Huang-Rhys theory------
\subsection{Photoluminescence spectrum within Huang-Rhys theory}
The total Huang-Rhys factor (HR-factor) is defined as the number of effective phonons participating in the optical transition which is a key parameter of the absorption and fluorescence spectra. The total HR-factor can be readily calculated within Franck-Condon approximation which assumes that the vibrational modes in the ground and excited states are identical. The associated phonon overlap spectral function can be derived from the overlap between the phonon modes in the electronic ground and excited states~\citep{alkauskas2014first, PhysRevLett.103.186404}. This phonon overlap spectral function can be employed to generate the phonon sideband of the photoluminescence spectrum. The details about the simulation of the photoluminescence spectrum can be found in the Appendix~\ref{sec:appendix A}.
%-end---Huang-Rhys theory------
%====================================================

%-INTERSTITIAL CLUSTERS------------------------------
\section{Results}
\label{sec:result}
\subsection{Interstitial clusters}
The carbon interstitial clusters are the simplest carbon aggregates observed in electron irradiated and implanted 4H-SiC samples~\citep{xu2009positron,yang2021anisotropic}. Some of these clusters can exhibit photoluminescence owing to their deep defect levels inside the band gap. In Secs.~\ref{sssec:Ci}-\ref{sssec:Ctetra}, we systematically discuss the local geometry, formation energy, dissociation energy, vibrational properties, fluorescence spectrum with ZPL energy and phonon sideband for carbon interstitial clusters.

%-ini---Carbon split interstitial------
\subsubsection{Carbon split-interstitial}
\label{sssec:Ci}
\begin{figure}[htb]
\includegraphics[width=0.4\textwidth]{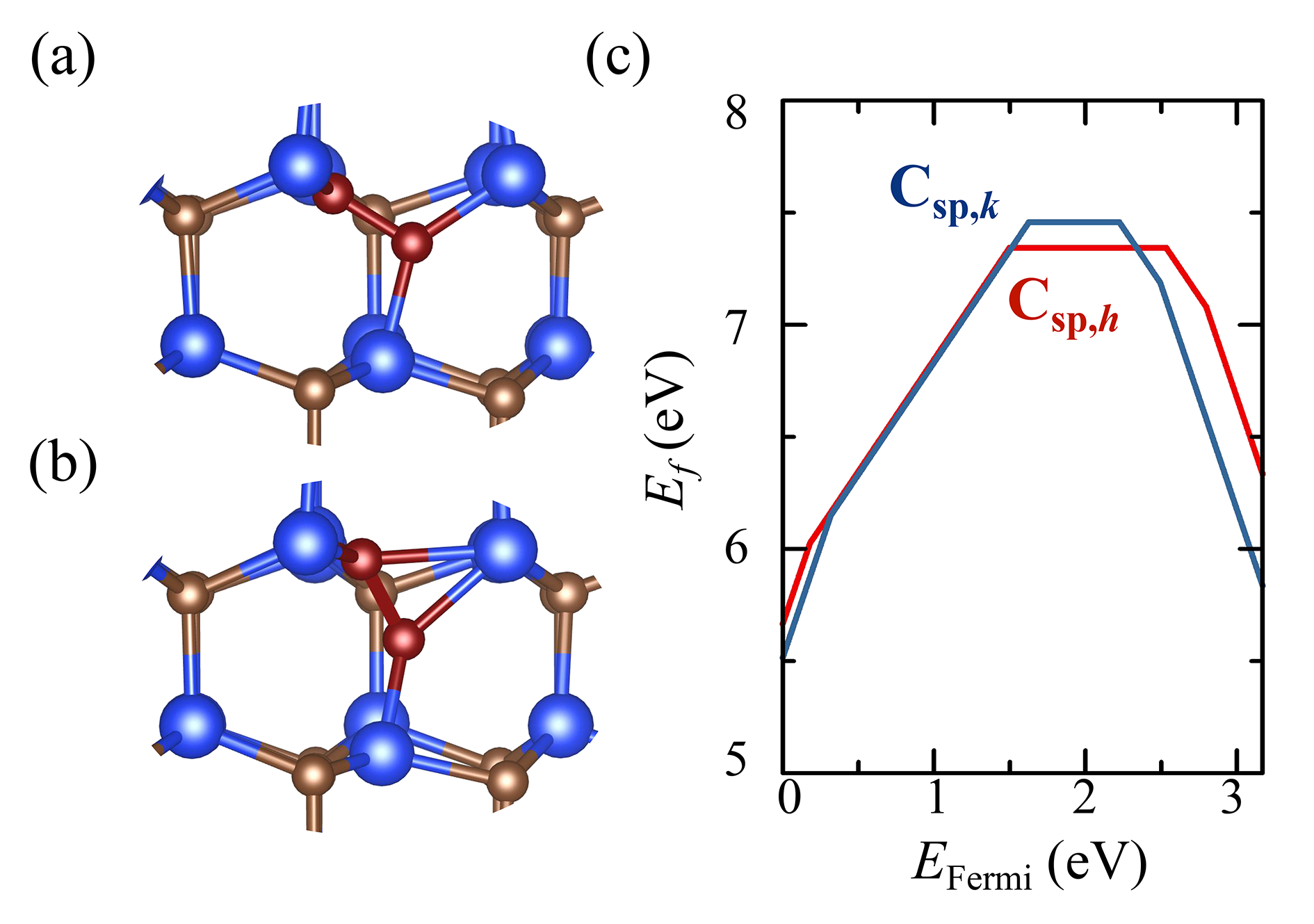}
\caption{\label{Figure 1}%
The local configurations of carbon split-interstitial ($\text{C}_\text{sp}$) at $h$ site: (a) $\text{C}_\text{sp}$ and (b) tilted $\text{C}_\text{sp}$. (c) Formation energy ($E_f$) of $\text{C}_\text{sp}$ as a function of the Fermi-level ($E_\text{Fermi}$) under carbon-rich condition.}
\end{figure}

Carbon split-interstitial ($\text{C}_\text{sp}$) is formed by trapping a carbon interstitial ($\text{C}_\text{i}$) on carbon lattice site~\citep{mattausch2004carbon}. Noteworthy, $\text{C}_\text{sp}$ is assumed to be a possible candidate for the \textit{P-T} centers in 6H-SiC by comparing the calculated and observed LVMs in the phonon sideband~\citep{mattausch2004structure, evans2002identification}. Hence, we speculate the $\text{C}_\text{sp}$ could also be a potential single photon source in 4H-SiC, especially the 463-nm triplet emitter which might have the same origin as the \textit{P-T} centers in 6H-SiC.     
Carbon split interstitials possess two dangling $p$-orbitals resulting from the sp$^2$ hybridization of the carbon atoms. The two adjacent split interstitials form a covalent bond due to the overlap of these orbitals. Compared with other interstitial carbon configurations, $\text{C}_\text{sp}$ has the lowest formation energy. Two inequivalent lattice sites, $h$ and $k$ sites, are considered here. At $h$ site [Fig.~\ref{Figure 1}(a)], $\text{C}_\text{sp}$ has $\text{S}_4$ symmetry and the bond lengths of C-C and Si-C are 1.36~\AA\ and 1.78~\AA, respectively. The formation energy of high-spin ($S=1$) $\text{C}_\text{sp}$ is 7.34~eV, 0.35~eV lower than that of the low-spin ($S=0$) defect state, indicating that the high-spin state is the more stable one. The tilted configuration~\citep{PhysRevB.65.184108} is nonmagnetic [see Fig.~\ref{Figure 1}(b)]. Within this configuration, the upper carbon atom moves upward and creates a bond with the adjacent silicon atom. The bond length of C-C is 1.34~\AA. The bond lengths of the nearest-neighbor Si-C are in the range of 1.72$\sim$2.07~\AA. However, the formation energy is 8.11~eV, thus it is a metastable form. At $k$ site, the formation energy of the high-spin, low-spin and tilted $\text{C}_\text{sp}$ are 7.46, 7.84 and 7.85~eV, respectively. The high-spin $\text{C}_\text{sp}$ is the most stable and tilted $\text{C}_\text{sp}$ with $S=0$ state has the highest energy, which is similar to the results at $h$ site. 

The formation energies of high-spin $\text{C}_\text{sp}$ are shown in Fig.~\ref{Figure 1}(c). For $\text{C}_{\text{sp}.h}$, the charge transition levels $\epsilon^{-2/-1}$ and $\epsilon^{-1/0}$ are $E_\text{C}-0.37$ and $E_\text{C}-0.64$ eV, respectively [$E_\text{C}$ is the energy of the conduction band minimum (CBM)]. These values agree well with the $\text{M}_1$ ($E_\text{C}-0.42$ eV) and $\text{M}_3$ ($E_\text{C}-0.75$ eV) from the deep-level transient spectroscopy (DLTS) spectrum `A' of the \textit{M} center in 4H-SiC~\cite{martin2004bistable}. For $\text{C}_{\text{sp}.k}$, $\epsilon^{-2/-1}$ and $\epsilon^{-1/0}$ are $E_\text{C}-0.68$ and $E_\text{C}-0.95$ eV, respectively. The former agrees well with $\text{M}_2$ peak in the DLTS spectrum `B' ($E_\text{C}-0.70$ eV) of \textit{M} center~\cite{martin2004bistable}. These results indicate that the $\text{C}_\text{sp}$ is responsible for the \textit{M} DLTS center in 4H-SiC, which is consistent with the conclusion achieved in Ref.~\onlinecite{coutinho2021m}.

The excitation of $\text{C}_\text{sp}$ can be described as promoting an electron from the highest occupied DL to the CBM. The calculated ZPL energies of $h$ and $k$ sites are 0.43~eV and 0.45~eV, respectively. These calculated values are much smaller than experimental data of optical centers in 4H-SiC (2$\sim$3~eV)~\citep{steeds2008creation}, excluding the possibility of $\text{C}_\text{sp}$ as a candidate for the observed PL centers.

%%%%% figure 1 %%%%
%%%%% figure 1 %%%%
%-end---Carbon split interstitial------

%-ini---di-carbon interstitial------
\subsubsection{Di-carbon interstitial clusters}
\begin{figure*}[htb]
\includegraphics[width=1.6\columnwidth]{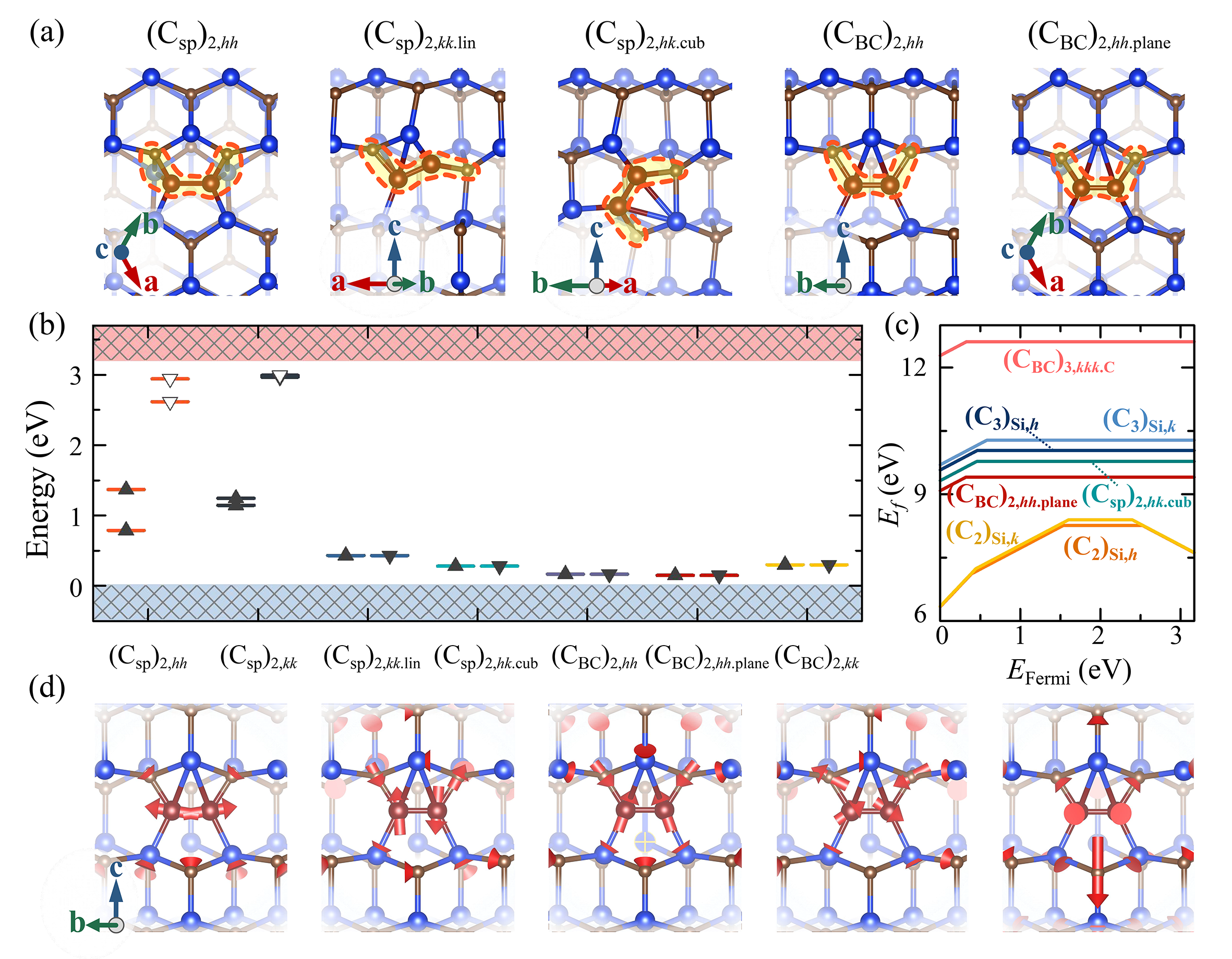}
\caption{\label{Figure 2}% 
(a) The local configurations of $(\text{C}_\text{sp})_{2,hh}$, $(\text{C}_\text{sp})_{2,kk.\text{lin}}$, $(\text{C}_\text{sp})_{2,hk.\text{cub}}$, $(\text{C}_\text{BC})_{2,hh}$ and $(\text{C}_\text{BC})_{2,hh.\text{plane}}$. $(\text{C}_\text{sp})_{2,kk}$ shares the same defect configuration with $(\text{C}_\text{sp})_{2,hh}$ at $k$ site. The same applies to the case of $(\text{C}_\text{BC})_{2,kk}$. (b) Calculated Kohn-Sham defect levels for the neutral di-carbon interstitials in the fundamental band gap of 4H-SiC. The occupied and unoccupied defect levels are labeled by filled and empty triangles, respectively. (c) Formation energy ($E_f$) as a function of the Fermi-level ($E_\text{Fermi}$) under carbon-rich condition. Di-carbon antisites, $(\text{C}_2)_\text{Si}$, are the most stable clusters. (d) The five vibration modes with the highest frequency of $(\text{C}_\text{BC})_{2,hh}$. The carbon interstitial atoms are depicted by red balls, and the respective vibrations are indicated by red arrows.}
\end{figure*}

Due to the unsaturated and reactive $p$ orbitals of the carbon interstitials, the aggregation of the two adjacent carbon interstitials forming di-carbon interstitial is a favorable reaction. The configuration of di-carbon interstitial is very complex as each of the self-interstitial can be at $h$ or $k$ site. In this section, we considered seven different geometries with $\text{C}_\text{sp}$ and $\text{C}_\text{BC}$ (the carbon interstitial occupying the Si-C bond-center site) shown in Fig.~\ref{Figure 2}(a). For the $\text{C}_\text{sp}$, we considered four di-carbon interstitial clusters: $(\text{C}_\text{sp})_{2,hh}$, $(\text{C}_\text{sp})_{2,kk}$, $(\text{C}_\text{sp})_{2,kk.\text{lin}}$ and $(\text{C}_\text{sp})_{2,hk.\text{cub}}$. For the $\text{C}_\text{BC}$, we considered three di-carbon interstitial clusters: $(\text{C}_\text{BC})_{2,hh}$, $(\text{C}_\text{BC})_{2,hh.\text{plane}}$ and $(\text{C}_\text{BC})_{2,kk}$. The results of the formation energy, dissociation energy, ZPL, HR-factor and LVMs are listed in Table~\ref{Table 1}.

%%%% figure 2 %%%%
%%%% figure 2 %%%%

%%%% table 1 %%%%
\begin{table*}[htb]
\caption{\label{Table 1} Formation energy ($E_f$), dissociation energy ($E_D$), ZPL ($E_\text{ZPL}$), HR-factor and LVMs in the neutral charge state of seven di-carbon interstitial clusters: $(\text{C}_\text{sp})_{2,hh}$, $(\text{C}_\text{sp})_{2,kk}$, $(\text{C}_\text{sp})_{2,kk.\text{lin}}$, $(\text{C}_\text{sp})_{2,hk.\text{cub}}$, $(\text{C}_\text{BC})_{2,hh}$, $(\text{C}_\text{BC})_{2,hh.\text{plane}}$ and $(\text{C}_\text{BC})_{2,kk}$. $\text{C}_\text{sp}$ is the carbon split-interstitial and $\text{C}_\text{BC}$ is the carbon interstitial occupying the Si-C bond-center site. $h$ and $k$ labels indicate the location of the interstitials.}
\begin{ruledtabular}
\begin{tabular}{c|cccc|ccccc}
di-carbon interstitial clusters & $E_f (\text{eV})$ & $E_D (\text{eV})$ & $E_\text{ZPL} (\text{eV})$ & HR-factor & \multicolumn{5}{c}{LVM (meV)} \\
\colrule\rule{0pt}{2.5ex}%
$(\text{C}_\text{sp})_{2,hh}$ & 11.94 & 2.74 & 1.18 & 4.6 & 119.9 & 126.2 & 127.6 & 159.9 & 161.8 \\  
$(\text{C}_\text{sp})_{2,kk}$ & 11.95 & 2.97 & 1.56 & 8.3 & 122.0 & 123.2 & 125.9 & 159.4 & 160.5 \\  
$(\text{C}_\text{sp})_{2,kk.\text{lin}}$ & 10.71 & 4.21 & 2.13 & 5.1 & 118.1 & 120.1 & 155.7 & 161.0 & 192.7 \\  
$(\text{C}_\text{sp})_{2,hk.\text{cub}}$ & 9.78 & 5.02 & 2.61 & 3.3 & 129.6 & 129.8 & 157.4 & 160.8 & 178.1 \\\hline  
$(\text{C}_\text{BC})_{2,hh}$ & 9.91 & 4.77 & 2.56 & 4.6 & 128.0 & 129.9 & 157.0 & 160.9 & 177.2 \\  
$(\text{C}_\text{BC})_{2,hh.\text{plane}}$ & 9.40 & 5.28 & 2.70 & 3.4 & 130.5 & 132.7 & 155.4 & 158.2 & 179.9 \\   
$(\text{C}_\text{BC})_{2,kk}$ & 9.85 & 5.07 & 2.59 & 2.9 & 129.2 & 129.6 & 156.9 & 160.2 & 179.6 \\   
\end{tabular}
\end{ruledtabular}
\end{table*}
%%%% table 1 %%%%

For the configuration $(\text{C}_\text{sp})_2$, two of the split interstitials form a long bond (1.48~\AA) by a rotation of the interstitial dumbbells towards each other [Fig.~\ref{Figure 2}(a)]. The difference of formation energies at $h$ and $k$ sites is 0.01~eV and the difference of the dissociation energies is 0.23~eV, which means that the first nearest neighbor configurations of $h$ and $k$ sites have almost no effect on the stability of the defects. The more stable configurations are $(\text{C}_\text{sp})_{2,kk.\text{lin}}$ and $(\text{C}_\text{sp})_{2,hk.\text{cub}}$ with dissociation energies of 4.21~eV and 5.02~eV, respectively. Compared with $\text{C}_\text{sp}$-related di-carbon interstitial clusters, $\text{C}_\text{BC}$-related di-carbon interstitial clusters are more stable, especially the $(\text{C}_\text{BC})_{2,hh.\text{plane}}$ which is the most stable. Thus, the carbon atoms prefer to occupy the Si-C bond-center sites, in contrast to the case of single carbon interstitial defects.

For the analysis of their electronic structure, we plot the in-gap Kohn-Sham levels of the neutral di-carbon interstitials in Fig.~\ref{Figure 2}(b). The neutral $(\text{C}_\text{sp})_{2,hh}$ and $(\text{C}_\text{sp})_{2,kk}$ are high-spin defects with two occupied spin-up DLs ($E_\text{V}$+0.79 and $E_\text{V}$+1.37~eV for $h$ site, $E_\text{V}$+1.15 and $E_\text{V}$+1.24~eV for $k$ site) and two unoccupied spin-down DLs ($E_\text{V}$+2.62 and $E_\text{V}$+2.94~eV for $h$ site, $E_\text{V}$+2.97 and $E_\text{V}$+2.99~eV for $k$ site). Other configurations have singlet ground state ($S=0$) and the defect levels are very near to the VBM ($E_\text{V}$+0.15$\sim$$E_\text{V}$+0.43~eV). By taking $(\text{C}_\text{sp})_{2,hk.\text{cub}}$ and $(\text{C}_\text{BC})_{2,hh.\text{plane}}$ as examples, the charge transition levels are shown in Fig.~\ref{Figure 2}(c), and the corresponding values of $\epsilon^{+/0}$ are $E_\text{V}$+0.45 and $E_\text{V}$+0.32~eV, respectively. Therefore, the neutral charge state is stable for a wide range of $E_\text{Fermi}$.

In order to identify the PL centers in experiments, we calculated the LVMs and the results are shown in Table~\ref{Table 1}. Here, to directly compare those with the experimental results, we only consider the vibration modes with the frequency larger than that of 4H-SiC bulk phonon spectrum (115.0~meV). For $(\text{C}_\text{sp})_{2,hh}$ and $(\text{C}_\text{sp})_{2,kk}$, the features in the spectrum are similar: two relatively close high-frequency LVMs [161.8 and 159.9~meV for $(\text{C}_\text{sp})_{2,hh}$, 160.5 and 159.4~meV for $(\text{C}_\text{sp})_{2,kk}$] and three LVMs close to the highest-frequency SiC bulk phonon mode (Table~\ref{Table 1}). The two highest-frequency modes are attributed to the stretching vibrations along the dumbbell formed by the interstitial and nearest neighbor carbon atom. The low-energy modes involve vibrations against the nearest neighbor atoms of the defect. For $(\text{C}_\text{sp})_{2,kk.\text{lin}}$, there are one high-frequency LVM (192.7~meV), two close frequency LVMs and two LVMs close to the highest SiC bulk phonon mode. The high-frequency modes originate from the vibrations of the carbon interstitial atoms, except for the fourth which originates from the vibrations of the first nearest neighbor carbon atoms. The features in the spectrum of $(\text{C}_\text{sp})_{2,hk.\text{cub}}$ and the three $\text{C}_\text{BC}$-related di-carbon interstitial clusters [$(\text{C}_\text{BC})_{2,hh}$, $(\text{C}_\text{BC})_{2,hh.\text{plane}}$ and $(\text{C}_\text{BC})_{2,kk}$] are similar: one highest-frequency, two close-frequency and two low-frequency LVMs (Table~\ref{Table 1}). The vibration modes are shown in Fig.~\ref{Figure 2}(d) by taking $(\text{C}_\text{BC})_{2,hh}$ as an example. The first highest-frequency mode corresponds to the C-C stretching mode roughly along the two carbon interstitials which preserves the symmetry of the defect. The second and third ones belong to the asymmetric and symmetric stretching vibrations of the interstitial and nearest neighbor carbon atoms. The fourth one is attributed to the asymmetric in-plane bending modes and the fifth one is from the axial vibration of the second nearest neighboring carbon atoms.

The ZPL energy ($E_\text{ZPL}$), depending on the electronic structure and ionic relaxation induced by excitation, is another criterion for potentially identifying an experimentally isolated emitter. It has been calculated for all the neutral di-carbon interstitial clusters mentioned above and the results are shown in Table~\ref{Table 1}. The excitation of $(\text{C}_\text{sp})_{2,hh}$ and $(\text{C}_\text{sp})_{2,kk}$ is from the highest occupied DL to CBM [$E_\text{ZPL} = 1.18$~eV for $(\text{C}_\text{sp})_{2,hh}$ and $E_\text{ZPL} = 1.56$~eV for $(\text{C}_\text{sp})_{2,kk}$]. For others, the excitation promotes one of the paired electrons from the highest occupied DL to CBM and the values of ZPL can be found in Table~\ref{Table 1}. The ZPL energies of $(\text{C}_\text{sp})_{2,hk.\text{cub}}$, $(\text{C}_\text{BC})_{2,hh}$, $(\text{C}_\text{BC})_{2,hh.\text{plane}}$ and $(\text{C}_\text{BC})_{2,kk}$ are in the range of 2.56$\sim$2.70~eV, which is consistent with the experimental results of the 463-nm triplet emitters in 4H-SiC ($E_\text{ZPL} = 2.68$~eV)~\citep{steeds2008creation}.  

\begin{figure}[htb]
\includegraphics[width=\columnwidth]{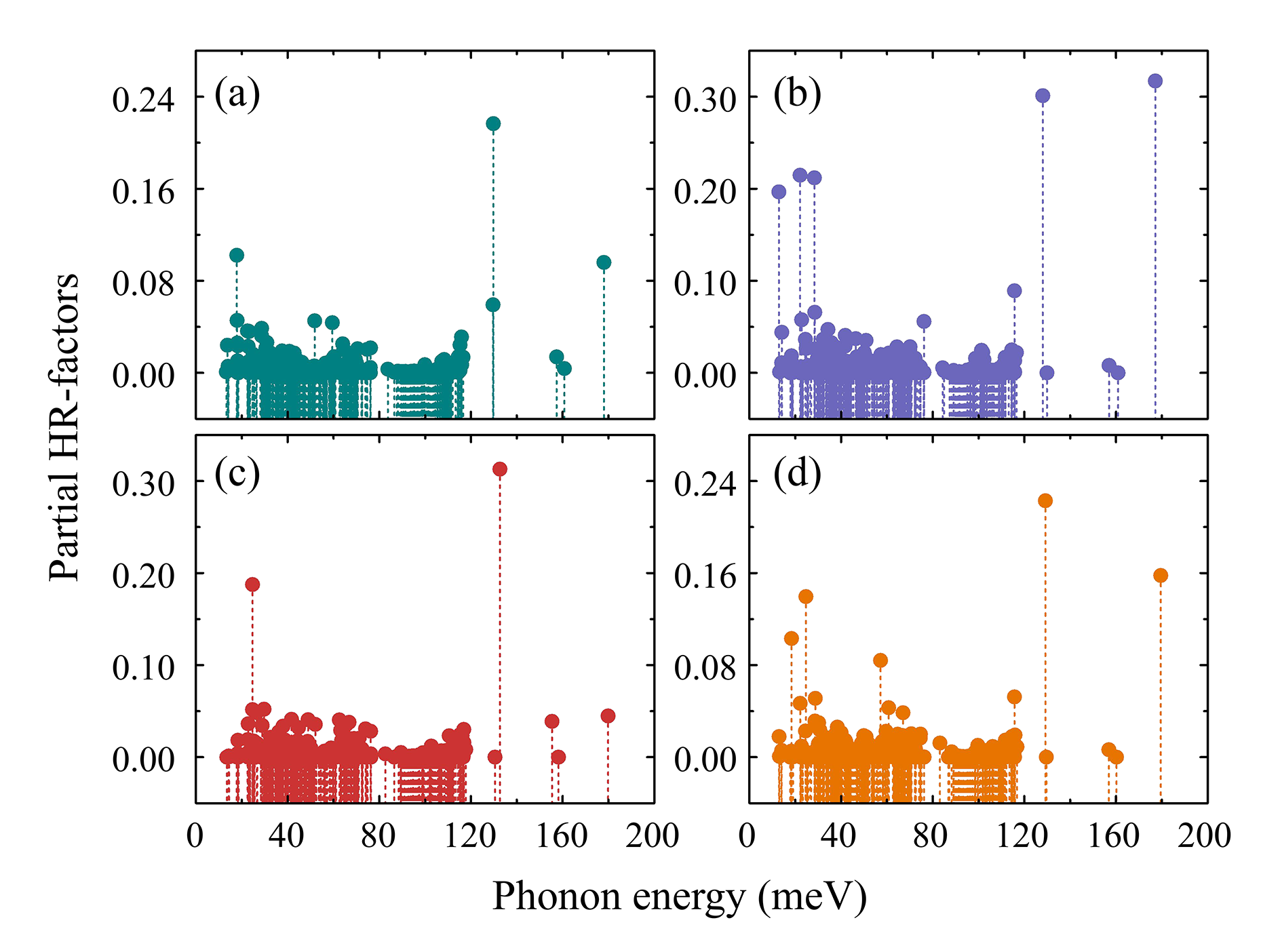}
\caption{\label{Figure 3}%
The partial HR-factors of (a) $(\text{C}_\text{sp})_{2,hk.\text{cub}}$, (b) $(\text{C}_\text{BC})_{2,hh}$, (c) $(\text{C}_\text{BC})_{2,hh.\text{plane}}$ and (d) $(\text{C}_\text{BC})_{2,kk}$.}
\end{figure}

To further support the identification of the 463-nm triplet emitters in 4H-SiC~\citep{steeds2008creation, steeds2008identification}, we calculated the partial HR-factors of $(\text{C}_\text{sp})_{2,hk.\text{cub}}$, $(\text{C}_\text{BC})_{2,hh}$, $(\text{C}_\text{BC})_{2,hh.\text{plane}}$ and $(\text{C}_\text{BC})_{2,kk}$. The results are shown in Fig.~\ref{Figure 3}. The HR-factors highlight the primary modes that participate in the excitation process. Except for $(\text{C}_\text{BC})_{2,hh.\text{plane}}$, the main contribution comes from the modes located at 177$\sim$180 and 129$\sim$133~meV (Fig.~\ref{Figure 3}). The results are in excellent agreement with the experimental data indicating that the $(\text{C}_\text{sp})_{2,hk.\text{cub}}$, $(\text{C}_\text{BC})_{2,hh}$ and $(\text{C}_\text{BC})_{2,kk}$ are responsible for the 463-nm triplet emitters in 4H-SiC. For $(\text{C}_\text{BC})_{2,hh.\text{plane}}$, the mode with the largest contribution is 132.7~meV [Fig.~\ref{Figure 3}(c)], which is consistent with the experimentally observed 456.7-nm emitter in 4H-SiC. 

%%%% figure 3 %%%%
%%%% figure 3 %%%%

\begin{figure}[htb]
\includegraphics[width=\columnwidth]{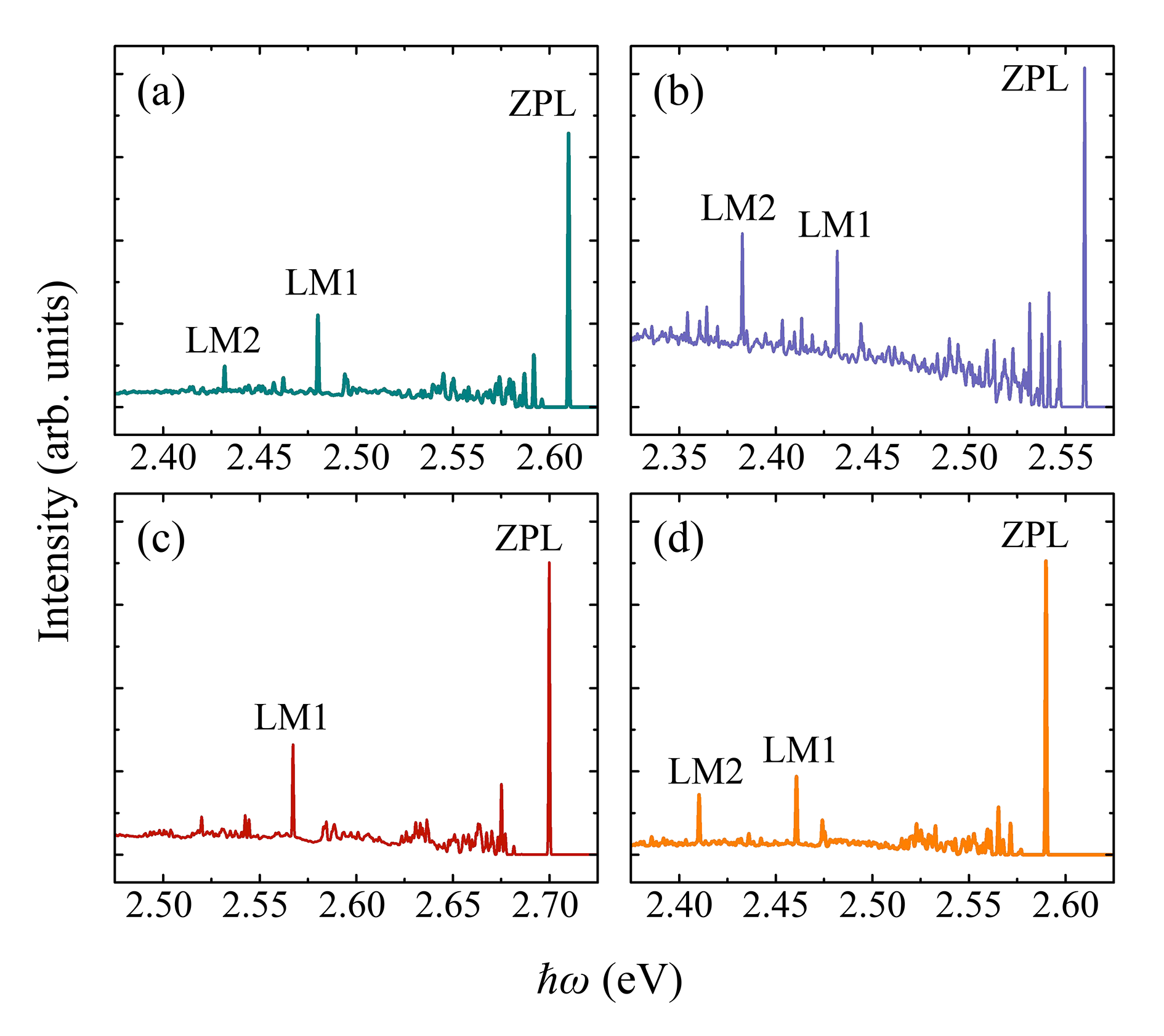}
\caption{\label{Figure 4}%
Calculated HR emission spectrum for the optical transition of (a) $(\text{C}_\text{sp})_{2,hk.\text{cub}}$, (b) $(\text{C}_\text{BC})_{2,hh}$, (c) $(\text{C}_\text{BC})_{2,hh.\text{plane}}$ and (d) $(\text{C}_\text{BC})_{2,kk}$. LM1 and LM2 are the effective LVMs during the excitation.}
\end{figure}

The calculated HR emission probability density functions of $(\text{C}_\text{sp})_{2,hk.\text{cub}}$, $(\text{C}_\text{BC})_{2,hh}$, $(\text{C}_\text{BC})_{2,hh.\text{plane}}$ and $(\text{C}_\text{BC})_{2,kk}$ are shown in Fig.~\ref{Figure 4}. 
%ZPLs in all spectra are shifted to the calculated ZPL positions as reported in Table~\ref{Table 2}. 
The continuous phonon sideband shifted with around 0.2$\sim$0.5 eV from the ZPL peak is dominated by low-energy collective phonon modes from SiC host. Compared to these broad and low intensity collective modes, we can clearly identify the sharp phonon peaks originated from local modes as labeled with LM1 and LM2. For $(\text{C}_\text{sp})_{2,hk.\text{cub}}$, $(\text{C}_\text{BC})_{2,hh}$ and $(\text{C}_\text{BC})_{2,kk}$, there are two outer LVMs (LM1 and LM2) which are related with the vibration of the defect. The higher-energy peak (LM1) is originated from the asymmetric in-plane bending mode or axial vibration of the second nearest neighboring carbon atoms. The other one (LM2) is originated from the C-C stretching mode. For $(\text{C}_\text{BC})_{2,hh.\text{plane}}$, there is only one outer LVM (LM1) from the asymmetric in-plane bending mode. The HR-factors of $(\text{C}_\text{sp})_{2,hk.\text{cub}}$, $(\text{C}_\text{BC})_{2,hh}$, $(\text{C}_\text{BC})_{2,hh.\text{plane}}$ and $(\text{C}_\text{BC})_{2,kk}$ in the range of 2.9$\sim$4.6 (Table~\ref{Table 1}). $(\text{C}_\text{BC})_{2,kk}$ has the smallest HR-factor of 2.9, indicating relatively weak electron-phonon coupling. 

%%%% figure 4 %%%%
%%%% figure 4 %%%%

Given the high mobility of carbon interstitials~\citep{bockstedte2003ab}, we analyzed the stability and the migration barrier taking the di-carbon interstitials at $h$ site [$(\text{C}_\text{sp})_{2,hh}$, $(\text{C}_\text{BC})_{2,hh}$ and $(\text{C}_\text{BC})_{2,hh.\text{plane}}$] as examples. The local configurations are shown in Fig.~\ref{Figure 2}(a). The most stable configuration is $(\text{C}_\text{BC})_{2,hh.\text{plane}}$, one pair of carbon interstitials lying between Si-C bond in the horizontal direction. The dissociation energy of $(\text{C}_\text{BC})_{2,hh}$ is 5.28~eV, higher than $(\text{C}_\text{BC})_{2,hh.\text{plane}}$ by 0.51~eV. $(\text{C}_\text{sp})_{2,hh}$ contains two neighboring $\text{C}_\text{sp}$ that are inclined towards each other and is the least stable among the three configurations with dissociation energy of 2.74~eV (2.8~eV in Ref.~\onlinecite{mattausch2004structure}). The migration barrier from $(\text{C}_\text{sp})_{2,hh}$ to $(\text{C}_\text{BC})_{2,hh}$ is 0.01~eV and it is 0.36~eV from $(\text{C}_\text{sp})_{2,hh}$ to $(\text{C}_\text{BC})_{2,hh.\text{plane}}$, which means $(\text{C}_\text{sp})_{2,hh}$ is metastable and can easily be transformed into $(\text{C}_\text{BC})_{2,hh}$ or $(\text{C}_\text{BC})_{2,hh.\text{plane}}$. The migration from $(\text{C}_\text{BC})_{2,hh}$ to $(\text{C}_\text{BC})_{2,hh.\text{plane}}$ is kinetically prohibited because of the large barrier energy (11.40~eV) resulting from the energy cost of breaking of strong covalent bonds. The results reaffirm that carbon interstitials in clusters have a tendency to occupy Si-C bond-center sites kinetically.
%-end---di-carbon interstitial------

%-ini---tri-carbon interstitial------
\subsubsection{Tri-carbon interstitial clusters}
\begin{figure}
\includegraphics[width=\columnwidth]{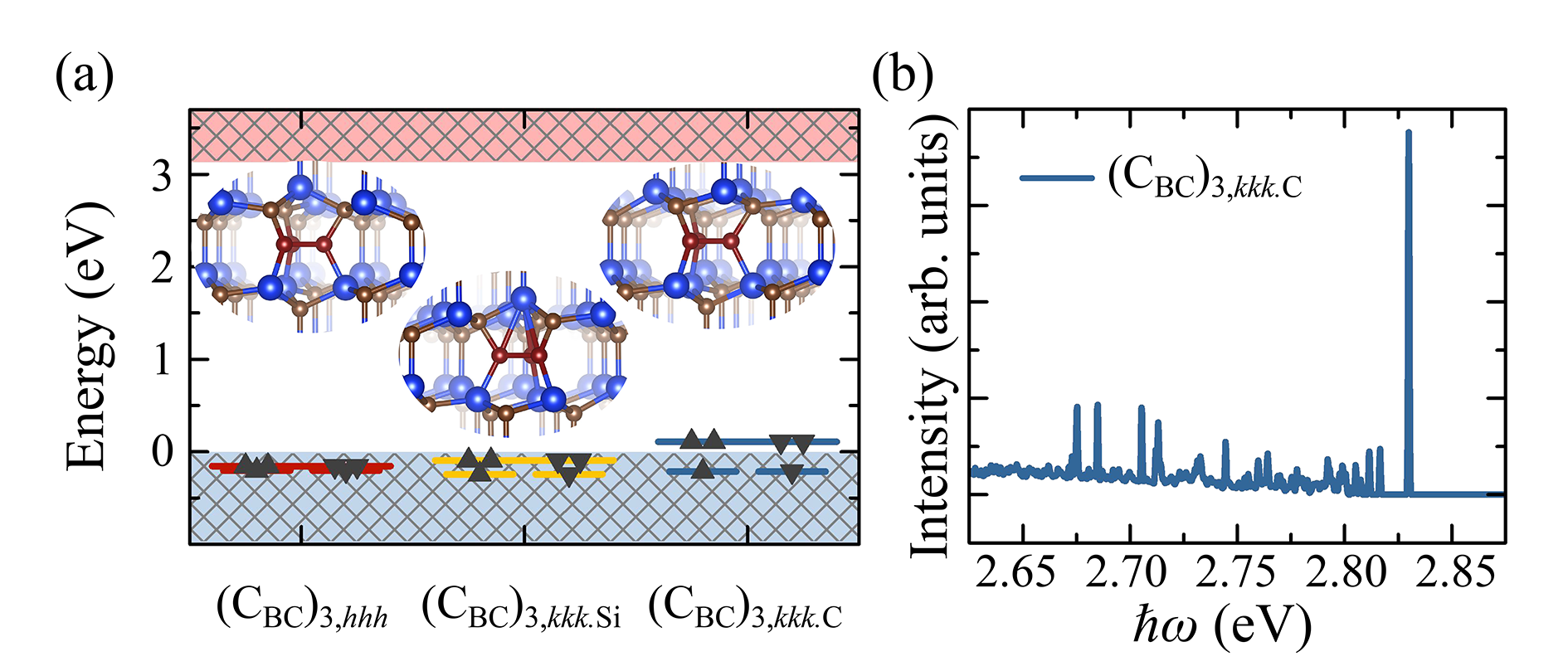}
\caption{\label{Figure 5}%
(a) Calculated Kohn-Sham defect levels for neutral tri-carbon interstitial clusters in the fundamental band gap of 4H-SiC. Insert: the configurations of $(\text{C}_\text{BC})_{3,hhh}$, $(\text{C}_\text{BC})_{3,kkk.\text{Si}}$ and $(\text{C}_\text{BC})_{3,kkk.\text{C}}$. $(\text{C}_\text{BC})_{3,kkk.\text{Si}}$ and $(\text{C}_\text{BC})_{3,kkk.\text{C}}$ refer to the cage where the lattice atom right above or below the center of the $(\text{C}_\text{BC})_{3,kkk}$ cluster are silicon or carbon atoms, respectively. (b) Calculated HR emission spectrum of the optical transition.}
\end{figure}

Because of the high mobility of carbon interstitials in 4H-SiC, $(\text{C}_\text{BC})_2$ can trap another interstitial carbon atom to form $(\text{C}_\text{BC})_3$ which has been proposed to be one of the most stable interstitial clusters in irradiated SiC~\citep{jiang2014structures, jiang2014accelerated}. This cluster is composed of three carbon interstitials occupying three neighboring Si-C bond center sites in the $[001]$ plane. Here, three diverse configurations are investigated, $(\text{C}_\text{BC})_{3,hhh}$, $(\text{C}_\text{BC})_{3,kkk.\text{Si}}$ and $(\text{C}_\text{BC})_{3,kkk.\text{C}}$ [Fig.~\ref{Figure 5}(a)]. The formation energy of the most stable $(\text{C}_\text{BC})_{3,hhh}$ is 12.02~eV and the dissociation energy is 5.23~eV. For $k$ site, $(\text{C}_\text{BC})_{3,kkk.\text{C}}$ is more stable than $(\text{C}_\text{BC})_{3,kkk.\text{Si}}$ by 0.55~eV, which is consistent with the result in Ref.~\onlinecite{jiang2014accelerated}.  $(\text{C}_\text{BC})_{3,hhh}$ and $(\text{C}_\text{BC})_{3,kkk.\text{Si}}$ are electrically inactive as they do not induce DLs inside the band gap [Fig.~\ref{Figure 5}(a)]. For $(\text{C}_\text{BC})_{3,kkk.\text{C}}$, there are four degenerate DLs with the value of $E_\text{V}$+0.11~eV. The charge transition levels of $(\text{C}_\text{BC})_{3,kkk.\text{C}}$ are shown in Fig.~\ref{Figure 2}(c), and the corresponding values of $\epsilon^{+/0}$ are $E_\text{V}$+0.33~eV, and the neutral state is stable for a wide range of $E_\text{Fermi}$.           

%%%% figure 5 %%%%
%%%% figure 5 %%%%

The LVMs of tri-carbon interstitial clusters are listed in Table ~\ref{Table 2}. $(\text{C}_\text{BC})_{3,hhh}$ has nine vibration modes with frequency higher than that of 4H-SiC bulk phonon spectrum. The two highest LVMs are stretching mode from two of the three approximately vertical C-C bonds. The third one includes the three C-C bond's stretching. The fourth one is a breathing mode of the triangle formed by the three carbon interstitials. The respective fifth and sixth ones result from the axial vibration of the carbon atoms right below and above the center of the $(\text{C}_\text{BC})_{3,hhhh}$ cluster. The sixth and seventh ones belong to the bending mode. The others are caused by the vibration of the neighbor carbon atoms. The vibration modes of $(\text{C}_\text{BC})_{3,kkk.\text{Si}}$ and $(\text{C}_\text{BC})_{3,kkk.\text{C}}$ are similar with that of $(\text{C}_\text{BC})_{3,hhh}$, except they only have one axial vibration [128.4~meV for $(\text{C}_\text{BC})_{3,kkk.\text{Si}}$ and 142.0~meV for $(\text{C}_\text{BC})_{3,kkk.\text{C}}$], which is determined by the arrangement of local geometry.

As $(\text{C}_\text{BC})_{3,hhh}$ and $(\text{C}_\text{BC})_{3,kkk.\text{Si}}$ are electrically inactive, we only consider the excitation of $(\text{C}_\text{BC})_{3,kkk.\text{C}}$. The value of ZPL is 2.83~eV\ and HR-factor is 3.53. The calculated HR emission spectrum for the optical transition is shown in Fig.~\ref{Figure 5}(b). There are four effective LVMs appearing in the fluorescence spectrum at 154.7, 145.1, 124.5 and 117.9~meV with reference to ZPL.       

%%%% table 2 %%%%
\begin{table}
\caption{\label{Table 2} Formation energy ($E_f$), dissociation energy ($E_D$), ZPL ($E_\text{ZPL}$), HR-factor and LVM in the neutral charge state of tri-carbon interstitial clusters. The $E_\text{ZPL}$ and HR-factor of $(\text{C}_\text{BC})_{3,hhh}$ and $(\text{C}_\text{BC})_{3,kkk.\text{Si}}$ are not calculated because of the lack of defect level in the band gap.}
\begin{ruledtabular}
\begin{tabular}{c|cccc}
defect & $(\text{C}_\text{BC})_{3,hhh}$ & $(\text{C}_\text{BC})_{3,kkk.\text{Si}}$ & $(\text{C}_\text{BC})_{3,kkk.\text{C}}$ \\
\colrule\rule{0pt}{2.5ex}%
$E_f (\text{eV})$ & 12.02 & 13.16 & 12.61 \\
$E_D (\text{eV})$ & 5.23 & 4.03 & 4.58 \\
$E_\text{ZPL} (\text{eV})$ &  &  & 2.83 \\
HR-factor &   &   & 3.53 \\\hline
LVM    & 117.1 & 118.1 & 117.2 \\
(meV)  & 126.4 & 118.1 & 117.3 \\
       & 126.4 & 121.7 & 117.9 \\
       & 128.6 & 125.1 & 124.5 \\
       & 141.7 & 125.2 & 124.6 \\
       & 146.1 & 128.4 & 142.0 \\
       & 157.6 & 148.1 & 145.1 \\
       & 162.0 & 156.5 & 154.7 \\
       & 162.0 & 158.2 & 156.9 \\
       &       & 158.3 & 156.9 \\
\end{tabular}
\end{ruledtabular}
\end{table}
%%%% table 2 %%%%

%-end---tri-carbon interstitial------

%-ini--tetra--carbon interstitial------
\subsubsection{Tetra-carbon interstitial clusters}
\label{sssec:Ctetra}
\begin{figure}[htb]
\includegraphics[width=\columnwidth]{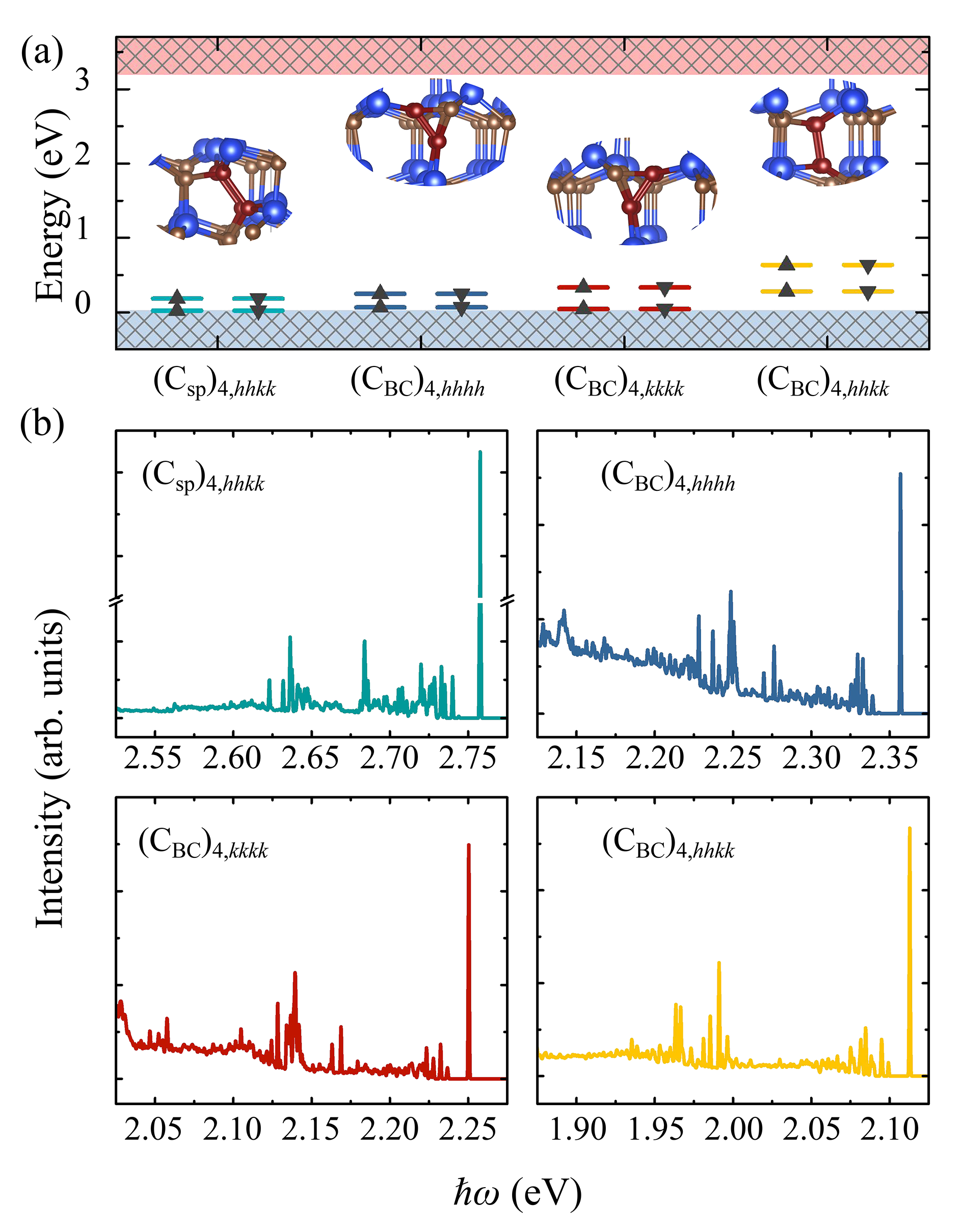}
\caption{\label{Figure 6}%
(a) Calculated Kohn-Sham defect levels for the neutral carbon tetra-carbon interstitial clusters in the fundamental band gap of 4H-SiC. Insert: the configurations of $(\text{C}_\text{sp})_{4,hhkk}$, $(\text{C}_\text{BC})_{4,hhhh}$, $(\text{C}_\text{BC})_{4,kkkk}$ and $(\text{C}_\text{BC})_{4,hhkk}$. (b) Calculated HR emission spectrum for the optical transition of the neutral tetra-carbon interstitials.}
\end{figure}

The tri-carbon interstitials can trap further carbon interstitials and form larger carbon clusters. Thus, we investigated the properties of the carbon clusters with four carbon interstitials. Here, we considered four different configurations: $(\text{C}_\text{sp})_{4,hhkk}$, $(\text{C}_\text{BC})_{4,hhhh}$, $(\text{C}_\text{BC})_{4,kkkk}$ and $(\text{C}_\text{BC})_{4,hhkk}$ [Fig.~\ref{Figure 6}(a)]. $(\text{C}_\text{sp})_{4,hhkk}$ is formed by four adjacent $\text{C}_\text{sp}$ defects (one pair on $h$ site and the other one on $k$ site). $(\text{C}_\text{BC})_{4,hhhh}$ and $(\text{C}_\text{BC})_{4,kkkk}$ are formed by four carbon interstitials occupying the Si-C bond-center sites (one pair along the horizontal direction and the other one along the adjacent vertical direction) and the four carbon atoms form a regular quadrilateral with a side length of 1.48~\AA. For $(\text{C}_\text{BC})_{4,hhkk}$, the four interstitials occupy the adjacent horizontal Si-C bonds. Among these configurations, $(\text{C}_\text{BC})_{4,hhkk}$ is the most stable with formation energy of 16.64~eV. The dissociation energy of $(\text{C}_\text{BC})_{4,hhhh}$ is negative ($-0.13$~eV), meaning the carbon atom does not really form. In Ref.~\onlinecite{mattausch2004structure}, $(\text{C}_\text{sp})_{4,hhkk}$ is regarded as the ideal model for tetra-carbon interstitial with the calculated dissociation energy of 5.3~eV. However, our results show that the formation energy of $(\text{C}_\text{BC})_{4,hhkk}$ is smaller than that of $(\text{C}_\text{sp})_{4,hhkk}$ by 1.05~eV. We conclude that the $(\text{C}_\text{BC})_{4,hhkk}$ should be the most stable configuration.  

%%%% figure 6 %%%%
%%%% figure 6 %%%%

%%%% table 3 %%%%
\begin{table}[htb]
\caption{\label{Table 3} Formation energy ($E_f$), dissociation energy ($E_D$), ZPL ($E_\text{ZPL}$), HR-factor and LVMs in the neutral charge state of tetra-carbon interstitial clusters.}
\begin{ruledtabular}
\begin{tabular}{c|cccc}
defect & $(\text{C}_\text{sp})_{4,hhkk}$ & $(\text{C}_\text{BC})_{4,hhhh}$ & $(\text{C}_\text{BC})_{4,kkkk}$ & $(\text{C}_\text{BC})_{4,hhkk}$ \\
\colrule\rule{0pt}{2.5ex}%
$E_f (\text{eV})$ & 17.69 & 19.49 & 19.58 & 16.64 \\
$E_D (\text{eV})$ & 2.94 & -0.13 & 1.05 & 2.72 \\
$E_\text{ZPL} (\text{eV})$ & 2.89 & 2.49 & 2.36 & 2.20 \\
HR-factor & 2.0 & 5.5 & 5.1 & 3.9 \\\hline
LVM & 118.4 & 118.8 & 116.4 & 119.8 \\
 (meV)   & 120.2 & 119.9 & 120.5 & 121.7 \\
    & 120.3 & 123.6 & 121.7 & 122.1 \\
    & 120.4 & 128.9 & 126.0 & 127.4 \\
    & 121.4 & 142.7 & 143.1 & 129.3 \\
    & 122.1 & 144.6 & 145.5 & 131.7 \\
    & 125.8 & 148.0 & 149.5 & 146.3 \\
    & 127.0 & 156.8 & 159.0 & 149.3 \\
    & 134.7 & 190.7 & 192.5 & 150.7 \\
    & 137.7 & 191.7 & 192.8 & 154.0 \\
    & 137.8 &       &       &       \\
    & 139.8 &       &       &       \\
\end{tabular}
\end{ruledtabular}
\end{table}
%%%% table 3 %%%%

The LVMs of the neutral tetra-carbon interstitial clusters are listed in Table ~\ref{Table 3}. For $(\text{C}_\text{BC})_{4,hhhh}$ and $(\text{C}_\text{BC})_{4,kkkk}$, the four higher modes are caused by the stretching vibration of the four base carbon atoms. The next two LVMs are caused by the vibration of the two apex carbon atoms. Others are due to the vibration of the second nearest neighboring carbon atoms. For $(\text{C}_\text{sp})_{4,hhkk}$, the four higher LVMs are attributed to the stretching vibrations along the dumbbell formed by the interstitial and nearest neighbor carbon atom, and the others presumably extend beyond the analyzed defect molecule. For $(\text{C}_\text{BC})_{4,hhkk}$, the two higher LVMs are caused by the vibration of the carbon interstitials. The 149.3 and 146.3~meV LVMs are due to the stretching mode of the horizontal C-C bonds. The fifth mode at 131.7~meV is resulted from the vibration of the second nearest neighboring carbon atoms. The next three modes are induced by the bending mode of the carbon atoms. Other modes are generated by the vibration of the second nearest neighbor carbon atoms.    

The four tetra-interstitial clusters have singlet ground state with producing four occupied DLs near VBM [Fig.~\ref{Figure 6}(a)]. The excitation promotes an electron from the highest occupied DL to CBM, and the values of ZPL are listed in Table~\ref{Table 3}. The calculated HR emission probability density functions are shown in Fig.~\ref{Figure 6}(b). The effective LVMs in the phonon sideband appear in the range of 120$\sim$150~meV near the ZPL peak. The smallest HR-factor at 2.0 corresponds to $(\text{C}_\text{sp})_{4,hhkk}$. For others, the HR-factor values are in the range between 3.9 to 5.5 (Table~\ref{Table 3}).

%====================================================

%-ANTISITE CLUSTERS------------------------------
\subsection{Antisite-related carbon clusters}
In addition to the carbon interstitial, we also analyzed the clusters formed by carbon antisites. In Secs.~\ref{sssec:CSi_1}-\ref{sssec:CSi_4}, we provide our calculated results of local geometry, formation energy, dissociation energy, ZPL energy and PL spectra for carbon antisites, di-, tri- and tetra-carbon antisite clusters.

%-ini---Carbon antisite------
\subsubsection{Carbon antisite}
\label{sssec:CSi_1}
For carbon point defects ($\text{V}_\text{C}$, $\text{C}_\text{i}$, $\text{C}_\text{Si}$), neutral carbon antisites ($\text{C}_\text{Si}$) possess the lowest formation energy among intrinsic point defects and are favorable under intrinsic and $n$-type conditions~\cite{kobayashi2019native, yan2020first}. The formation energies of the neutral $\text{C}_\text{Si}$ configurations are 4.68~eV and 4.64~eV at $h$ and $k$ sites, respectively. However, despite its high stability, theoretical calculations suggest that it is electrically inactive as they do not introduce deep levels inside the band gap. Our result is consistent with previous ones~\cite{MATTAUSCH2001656}. Therefore, it is difficult to optically or electrically detect in the experiment. Nevertheless, $\text{C}_\text{Si}$ is of a particular importance as a nucleus for larger carbon aggregates and can trap carbon interstitial atoms~\cite{gali2003aggregation, mattausch2004carbon}. When carbon interstitials are present in the SiC crystal lattice, they can interact with the carbon antisite defect and form complexes with short carbon-carbon bonds.
%-end---Carbon antisite------

%-ini---Carbon di-antisite------
\subsubsection{Di-carbon antisite clusters}
The di-carbon antisite defect, $(\text{C}_2)_\text{Si}$, is produced by pairing $\text{C}_\text{i}$ and a carbon antisite. The local structure is similar to that of $\text{C}_\text{sp}$, except the first nearest neighbor atoms are carbon and not silicon atoms. At $h$ site, the bond length of C-C and nearest neighbor C-C are 1.42 and 1.51~\AA, respectively. The formation energies of the triplet (high-spin), open-shell and closed-shell singlet (low-spin) configurations are 8.26, 8.25 and 8.83~eV, respectively. The formation energy of triplet is lower than that of closed-shell singlet by 0.57~eV and higher than that of open-shell singlet by 0.01~eV. We note that the total energy of the open-shell singlet might be less accurate than that of the triplet within spinpolarized Kohn-Sham hybrid density functional approach. To overcome this challenge and obtain accurate results, higher-level methods are required to address the spin contamination and ensure reliable calculations which is out of the scope of the present study. In view of the negligible energy difference (0.01~eV) between triplet and open-shell singlet, we think the triplet state can be observed at room temperature by electron paramagnetic resonance (EPR) spectroscopy. At $k$ site, the results are similar to that at $h$ site. And the formation energies of triplet (high-spin), open-shell and closed-shell singlet (low-spin) configurations are 8.40, 8.39 and 8.98~eV, respectively. In addition, the formation energy of tilted $(\text{C}_2)_\text{Si}$ are higher at both sites (9.28~eV for $h$ site and 9.99~eV for $k$ site) manifesting metastable configurations. Therefore, we only consider the high-spin model for $(\text{C}_2)_\text{Si}$ clusters at $h$ and $k$ sites in the following discussion. The formation energies and charge transition levels are shown in Fig.~\ref{Figure 2}(c). $(\text{C}_2)_{\text{Si},h}$ [$(\text{C}_2)_{\text{Si},k}$] is neutral when $E_\text{Fermi}$ is in the range of $E_\text{V}$+1.54$\sim$$E_\text{V}$+2.52 [$E_\text{V}$+1.60$\sim$$E_\text{V}$+2.40]~eV.      

%%%% table 4 %%%%
\begin{table}[htb]
\caption{\label{Table 4} Formation energy ($E_f$), dissociation energy ($E_D$), ZPL ($E_\text{ZPL}$) energy, HR-factor and LVMs in the neutral charge state of antisite-related carbon clusters.}
\begin{ruledtabular}
\begin{tabular}{c|cccccc}
defect & $(\text{C}_{2})_{\text{Si},h}$ & $(\text{C}_{2})_{\text{Si},k}$ & $(\text{C}_{3})_{\text{Si},h}$ & $(\text{C}_{3})_{\text{Si},k}$ & $(\text{C}_{4})_{\text{Si},h}$ & $(\text{C}_{4})_{\text{Si},k}$ \\
\colrule\rule{0pt}{2.5ex}%
$E_f (\text{eV})$ & 8.26 & 8.40 & 10.04 & 10.28 & 14.36 & 14.54 \\
$E_D (\text{eV})$ & 3.76 & 3.70 & 2.12 & 1.97 & -0.41 & -0.42 \\
$E_\text{ZPL} (\text{eV})$ & 1.60 & 1.54 & 2.67 & 2.56 & 1.82 & 1.51 \\
HR-factor & 2.91 & 3.09 & 2.35 & 2.59 & 33.15 & 13.15 \\\hline
LVM & 119.4 & 118.5 & 116.4 & 116.8 & 116.6 & 115.8 \\
(meV) & 130.3 & 130.9 & 121.7 & 121.8 & 116.7 & 115.8 \\
    & 131.7 & 132.1 & 126.9 & 126.7 & 116.8 & 115.9 \\
    & 170.9 & 170.5 & 150.7 & 153.2 & 116.8 & 116.0 \\
    &       &       & 173.9 & 174.6 & 116.9 & 116.0 \\
    &       &       & 246.7 & 248.6 & 192.0 & 193.0 \\
    &       &       &       &       & 193.5 & 193.0 \\
    &       &       &       &       & 193.5 & 193.7 \\
    &       &       &       &       & 231.7 & 233.0 \\
\end{tabular}
\end{ruledtabular}
\end{table}
%%%% table 4 %%%%

The LVMs of $(\text{C}_2)_\text{Si}$ are listed in Table~\ref{Table 4}. For $h$ site, the highest LVM is 170.9~meV, attributed to the stretching vibrations along the dumbbell formed by the antisite carbon atoms. The LVMs with the values of 131.7 and 130.3~meV are attributed to the bending mode of the two carbon atoms. The lowest LVM is also caused by the stretching mode of the defect. The $k$ site defect [$(\text{C}_2)_{\text{Si},k}$] shows similar results.

\begin{figure}[htb]
\includegraphics[width=\columnwidth]{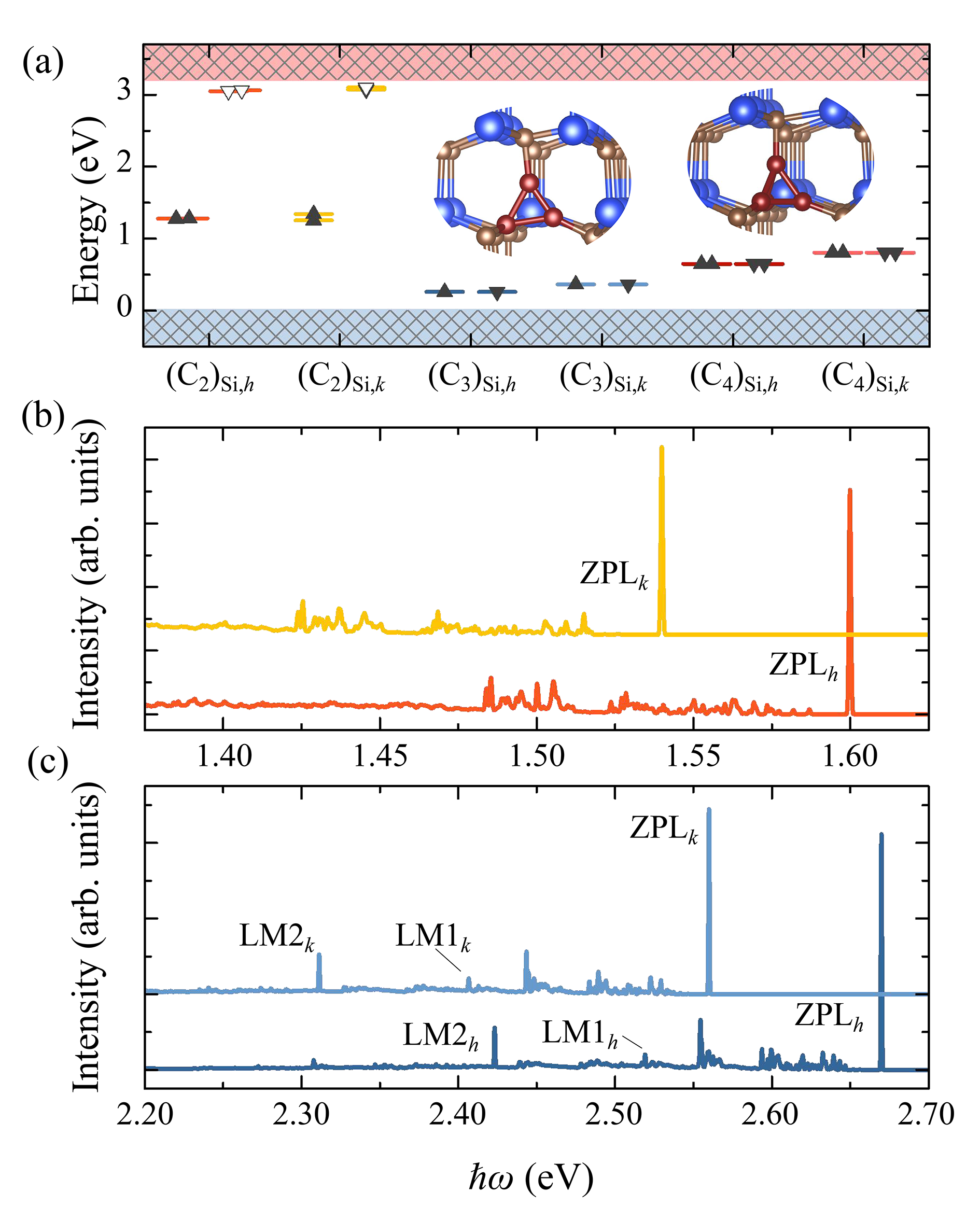}
\caption{\label{Figure 7}%
(a) Calculated Kohn-Sham defect levels for neutral di-, tri- and tetra-carbon antisite carbon clusters in the fundamental band gap of 4H-SiC. Inset: the configurations of $(\text{C}_3)_\text{Si}$ and $(\text{C}_4)_\text{Si}$ at $h$ site. The occupied and unoccupied defect levels are labeled by filled and empty triangles, respectively. (b) and (c) are calculated HR emission spectra for the optical transition of the neutral di- and tri-carbon antisite carbon clusters at $h$ and $k$ sites, respectively. The spectra are offset vertically for clarity.}
\end{figure}

The in-gap Kohn-Sham levels of the neutral defects are shown in Fig.~\ref{Figure 7}(a). The neutral $(\text{C}_2)_{\text{Si},h}$ and $(\text{C}_2)_{\text{Si},k}$ bears high-spin with two occupied spin-up DLs and two unoccupied spin-down DLs. The lowest excitation energies can be described as promoting an electron from DL to CBM at both $h$ and $k$ sites with ZPL energies at 1.60~eV and 1.54~eV, respectively (Table~\ref{Table 4}). The ZPL energies of di-carbon antisites fall close to the V1 (1.438~eV) and V2 (1.352~eV) PL centers in 4H-SiC~\citep{wagner2000electronic} which are identified as the negative silicon-vacancy at $h$ and $k$ sites based on various factors such as zero-field splitting, ZPL energies, and hyperfine constants of $^{13}$C and $^{29}$Si $I=1/2$ nuclear spins~\citep{ivady2017identification}.

The calculated fluorescence spectra are shown in Fig.~\ref{Figure 7}(b). No local vibration mode appears but broad and low intensity phonons belonging to the collective modes from the host. The local modes in Table~\ref{Table 4} do not participate to the fluorescence spectrum. This also happens to $(\text{C}_2)_{\text{Si},k}$. The calculated partial HR-factors of these defects are plotted in  Appendix~\ref{sec:appendix B} which clearly demonstrate our claim. The total HR-factors of $(\text{C}_2)_{\text{Si},h}$ and $(\text{C}_2)_{\text{Si},k}$ are 2.91 and 3.09, respectively. We note that only the neutral charge state of di-carbon antisites is considered in our calculation. It might have different emissions in the other charge states, such as the positive (negative) charge state in $p$-type($n$-type) 4H-SiC.

%%%% figure 7 %%%%
%%%% figure 7 %%%%

The paired $(\text{C}_2)_\text{Si}$, $[(\text{C}_2)_\text{Si}]_2$, is considered as the origin of the $D_{\text{II}}$ center in Ref.~\onlinecite{PhysRevB.69.235202} due to the good agreement in the values of LVM. The highest LVM of $[(\text{C}_2)_\text{Si}]_2$ is 159.6~meV, similar to that of $D_{\text{II}}$ center (164.1 meV in Ref.~\onlinecite{sullivan2007study}). The formation energy for the defect at $h$ site is 13.34~eV with dissociation energy of 1.62~eV, implying the $(\text{C}_2)_\text{Si}$ defects tend to bond with each other and form a larger cluster. However, the calculated ZPL is 1.46~eV, which is significantly smaller than that of $D_{\text{II}}$ center (3.205~eV in Ref.~\onlinecite{sullivan2007study}). Hence, $[(\text{C}_2)_\text{Si}]_2$ cannot be a reasonable candidate for $D_{\text{II}}$ center.
%-end---Carbon di-antisite------

%-ini---Carbon tri-antisite------
\subsubsection{Tri-carbon antisite clusters}
The di-carbon antisite may trap another carbon interstitial and form the tri-carbon antisite [$(\text{C}_3)_\text{Si}$]. The tri-carbon antisite is isosceles triangle-shaped defect with side lengths of 1.29, 1.47 and 1.47~\AA\ [Fig.~\ref{Figure 7}(a)]. Two of the three carbon atoms in the core of the defect have sp$^2$ hybridization, and the apex one is surrounded by four carbon atoms, forming sp$^3$ hybridization. Tri-carbon antisite is neutral when $E_\text{Fermi}$ lies higher than $E_\text{V}$+0.46 ($E_\text{V}$+0.58)~eV for $h$ ($k$) site [Fig.~\ref{Figure 2}(c)]. The formation energies of the neutral $(\text{C}_{3})_{\text{Si},h}$ and $(\text{C}_{3})_{\text{Si},k}$ are 10.04 and 10.28~eV (Table~\ref{Table 4}), respectively. The dissociation energies are 2.12 and 1.97~eV (Table~\ref{Table 4}), indicating that $(\text{C}_2)_\text{Si}$ prefers to capture an additional carbon interstitial atom and forms a more stable $(\text{C}_3)_\text{Si}$ cluster. The formation process and intermediate configurations have been investigated and confirmed in Ref.~\onlinecite{mattausch2004structure}. 

The tri-carbon antisite cluster shows a pronounced LVM spectrum (Table~\ref{Table 4}). The highest LVMs of $(\text{C}_{3})_{\text{Si},h}$ and $(\text{C}_{3})_{\text{Si},k}$ is located at 246.7 and 248.6~meV, respectively, which belong to the stretching vibrations between the two sp$^2$ carbon atoms. The second one with the value of 173.9 (174.6)~meV at $h$ ($k$) site is an asymmetric stretching vibration of the two sp$^2$ carbon atoms against their first nearest neighbors from the enclosing tetrahedron. The third one is the stretching of the apex carbon atom against the two sp$^2$ carbon atoms. The fourth one is due to the stretching vibration of two outer carbon atoms bonding with the sp$^2$ carbon atoms. The others are the vibration of the apex carbon atom along different directions. The results are consistent with previous calculations~\cite{mattausch2004structure}.

The tri-carbon antisite cluster is spin singlet and there are two occupied DLs close to VBM as shown in Fig.~\ref{Figure 7}(a): $E_\text{V}$+0.26~eV for $h$-site and $E_\text{V}$+0.36~eV for $k$-site. The excitation is from the highest occupied DL to CBM, and the values of ZPL are at 2.67 and 2.56~eV for $h$ and $k$ sites, respectively. The PL spectrum is shown in Fig.~\ref{Figure 7}(c). For both defects, $(\text{C}_{3})_{\text{Si},h}$ and $(\text{C}_{3})_{\text{Si},k}$, there are two effective LVM (LM1 and LM2) in the phonon sideband. The sharp phonon peak, LM1 ($\text{LM1}_h$ and $\text{LM1}_k$), is due to the stretching vibration along the sp$^2$ C-C bond. The low intensity one ($\text{LM2}_h$ and $\text{LM2}_k$) belongs to the stretching of the apex carbon atom against the two sp$^2$ atoms. The HR-factors of $(\text{C}_{3})_{\text{Si},h}$ and $(\text{C}_{3})_{\text{Si},k}$ are 2.35 and 2.59, respectively.

Our calculated results of $(\text{C}_3)_\text{Si}$ are similar to the experimental results of the 471.8-nm emitter (2.627~eV) in 4H-SiC~\citep{steeds2008identification}. The calculated ZPL energies of the $(\text{C}_{3})_{\text{Si},h}$ and $(\text{C}_{3})_{\text{Si},k}$ are at 2.67 and 2.56~eV. The calculated LVMs in Table~\ref{Table 4} are consistent with the experimental results in Ref.~\onlinecite{steeds2008identification}. However, the 173.9 (174.6) meV LVM for $(\text{C}_{3})_{\text{Si},h}$ [$(\text{C}_{3})_{\text{Si},k}$] has no contribution to the excitation as shown in Fig.~\ref{Figure 7}(c) as confirmed by the partial HR-factor calculations (see Appendix~\ref{sec:appendix B}). This result clearly shows that the calculated LVMs cannot be directly used to identify a PL center and the phonon sideband should be rather calculated at \emph{ab initio} level. We here assign one of the tri-carbon antisite defects as the origin of 471.8-nm emitter in 4H-SiC. Further investigation is still needed to confirm whether the feature at 189.4~meV in the observed PL spectrum indeed belongs to 471.8-nm emitter in 4H-SiC.   
%-end---Carbon tri-antisite------

%-ini---Carbon tetra-antisite------
\subsubsection{Tetra-carbon antisite clusters}
\label{sssec:CSi_4}
Tetra-carbon antisite cluster [$(\text{C}_4)_\text{Si}$] is formed by trapping one more carbon interstitial by $(\text{C}_3)_\text{Si}$. The local geometry at $h$ site is shown in Fig.~\ref{Figure 7}(a). For $h$ site, the four carbon atoms form a tetrahedron. Along the $\langle0001\rangle$ direction, the lengths of the base C-C bonds and edge C-C bonds are 1.45 and 1.46~\AA, respectively. For $k$ site, the four carbon atoms form a regular tetrahedron and the length of the edge C-C bonds is 1.45~\AA. At $h$ ($k$) site, the DLs are double degenerate states at $E_\text{V}$+0.65 ($E_\text{V}$+0.81)~eV. The formation energy of $(\text{C}_4)_\text{Si}$ is 14.36~eV for $h$ site and 14.54~eV for $k$ site. The calculated dissociation energy is negative ($-0.41$~eV for $h$ site and $-0.42$~eV for $k$ site), meaning that these clusters are not stable species. Nevertheless, for the sake of completeness, we show the calculated LVMs above the bulk phonon spectrum
in Table~\ref{Table 4}.
%====================================================

\section{Discussion}
\label{sec:discussion}
The accurate calculation of the electronic structure and photolumiescence spectrum may be enabled to firm identification the identification of previously reported carbon-cluster related PL centers in 4H-SiC. We note that many carbon-cluster related PL centers were reported in 6H-SiC too that might be common with those in 4H-SiC. Therefore, we list the reported PL centers in both polytypes together with the calculated ones in this study in Tables~\ref{Table 5} and~\ref{Table 6} with providing the appropriate references.

%%%% table 5 %%%%
\renewcommand{\multirowsetup}{\centering}
\begin{table*}[htb]
\caption{\label{Table 5} The experimental results of wavelengths ($\lambda_\text{ZPL}$) and energies ($E_\text{ZPL}$) of ZPLs together with the effective LVMs of the reported PL centers in 6H-SiC~\citep{evans2002identification} and 4H-SiC~\citep{steeds2008creation,steeds2008identification}.}
\begin{ruledtabular}
\begin{tabular}{c|ccc|c|c|c||c|c|ccc|c}
\multicolumn{6}{c}{\makecell{PL centers in 6H-SiC~\citep{evans2002identification}}} & & \multicolumn{6}{c}{\makecell{PL centers in 4H-SiC~\citep{steeds2008creation,steeds2008identification}}} \\\hline
& \multicolumn{2}{c}{\makecell{LVMs \\ (meV)}} & & \makecell{$\lambda_\text{ZPL}$ \\ (nm)} & \makecell{$E_\text{ZPL}$ \\ (eV)}  & \makecell{$E_\text{ZPL}+0.23$ \\ (eV)} & \makecell{$E_\text{ZPL}$ \\ (eV)} & \makecell{$\lambda_\text{ZPL}$ \\ (nm)} &  \multicolumn{2}{c}{\makecell{LVMs \\ (meV)}} &  &\\ \hline
\textit{P} & 133.2 & 179.3  &        & 502.3 & 2.47 &  2.70    &       &      &        &        &    & \\
\textit{Q} & 132.7 & 178.1  &        & 510.1 & 2.43 &  2.66    & 2.68  & 463.3 & 132.82 & 179.86 &    & \multirow{3}*{\makecell{triplet \\ emitter }}  \\
\textit{R} & 132.3 & 180.2  &        & 510.8 & 2.43 &  2.66    & 2.67  & 463.6 & 132.53 & 178.47 &       & \\
\textit{S} & 133.0 & 178.9  &        & 513.4 & 2.41 &  2.64    & 2.67  & 464.3 & 131.90 & 180.03 &       & \\
\textit{T} & 132.9 & 178.4  &        & 514.6 & 2.41 &  2.64    &       &       &        &        &       & \\ \hline
\textit{U} & 132.9 & 246.6  &        & 525.2 & 2.36 &  2.59    & 2.63  & 471.8 & 151.8  & 189.4  & 246.9 & \\
\textit{Z} & 119.2 & 173.3  & 203.3  & 512.9 & 2.42 &  2.65    &       &       &        &        &       & \\ 
\end{tabular}
\end{ruledtabular}
\end{table*}
%%%% table 5 %%%%

We learnt that many stable carbon clusters produce deep levels close to the valence band with establishing a closed-shell singlet state. The excitation of these defects can be described as promoting an electron from the deep defect level to CBM. There are two serious consequences of this observation: (i) the excited state is a bound exciton where the hole is strongly localized and the electron is Coulombically bound by the hole, so a Rydberg series of excited states should exist with following the effective mass state theory; (ii) the ZPL energy of the common defects in 4H-SiC and 6H-SiC should follow the energy difference of CBM edges in the two materials, i.e., they should be about 0.23~eV higher in 4H-SiC than those in 6H-SiC. We note that the $D_{\text{I}}$ centers in these polytypes follow this pattern~\cite{gali2003correlation, haberstroh1994some} for which the bound exciton nature of the excited state was proven by photoluminescence excitation (PLE) experiments~\cite{egilsson1999properties, patrick1972photoluminescence, haberstroh1994some}.

By comparing the $E_\text{ZPL}$ and LVMs in 4H-SiC and 6H-SiC, we find that \textit{P-T} centers in 6H-SiC (that form always simultaneously) correspond to the so-called triplet emitter in 4H-SiC (see Table~\ref{Table 5}). Our HR-theory confirms that LVMs at around 130~meV and 180~meV should appear in the PL spectra from three configurations di-carbon interstitial clusters in 4H-SiC, and their calculated $E_\text{ZPL}$ are in good agreement with the experimental data. These strongly suggest that the origin of the triplet-emitters in 4H-SiC and the \textit{P-T} PL centers in 6H-SiC are the stable di-carbon interstitial clusters. We note that the di-carbon antisite clusters were previously assigned to these PL centers based on the calculated LVMs~\cite{gali2003aggregation, mattausch2004carbon}. Indeed, LVMs can be found at around 130~meV and 170~meV for these clusters (see Table~\ref{Table 4}), not far from the features in the experiments. However, these LVMs are not optically active, and the calculated $E_\text{ZPL}$ in the near-infrared together with the optically active LVM at 114~meV do not agree at all with the observations (see Table~\ref{Table 6}). This issue again demonstrates the need of direct comparison of the simulated and observed PL spectra. 

According to annealing studies, the \textit{P-T} centers in 6H-SiC and the triplet-emitter in 4H-SiC start to dissociate at 900~$^{\circ}$C~\cite{evans2002identification,steeds2008identification}. Our results then imply that the di-carbon interstitial clusters dissociates that will re-emit mobile carbon interstitials to form larger and thermally more stable aggregates. Another more stable form of di-carbon interstitial defect [$(\text{C}_\text{BC})_{2,hh.\text{plane}}$] in 4H-SiC only produces a single LVM at $\sim$133~meV in the PL spectrum with $E_\text{ZPL} \approx 2.7$~eV, which has a similar optical characteristic to the 456.7-nm emitter in 4H-SiC (Table~\ref{Table 6}). The 456.7-nm emitter displays distinct annealing characteristics compared to the triplet-emitter. At a temperature of 1200~$^{\circ}$C, the former maintains a high density while the latter nearly disappears~\cite{steeds2008identification}. Interestingly, no similar PL center was observed in 6H-SiC in terms of single LVM. 

%%%% table 6 %%%%
\renewcommand{\multirowsetup}{\centering}
\begin{table*}[htb]
\caption{\label{Table 6} The comparison of wavelengths ($\lambda_\text{ZPL}$) and energies ($E_\text{ZPL}$) of ZPLs together with their effective LVMs during the excitation between the previously reported PL centers and our calculated results of carbon clusters in 4H-SiC. For di-carbon antisite clusters, the frequency of the highest effective LVM is lower than that of the 4H-SiC bulk phonon spectrum (115.0 meV) and the higher LVMs (list in Table~\ref{Table 4}) have no contribution to the excitation. For the optical centers in $\text{4H-SiC}/\text{SiO}_2$ interface, we only show the ranges of LVMs.}
\begin{ruledtabular}
\begin{tabular}{c|c|c|ccc||c|c|ccccc}
\multicolumn{5}{c}{\makecell{experimental results~\citep{steeds2008creation,steeds2008identification,johnson2019optically}}} & & \multicolumn{7}{c}{\makecell{our calculated results}} \\\hline
  &\makecell{$\lambda_\text{ZPL}$ \\ (nm)} & \makecell{$E_\text{ZPL}$ \\ (eV)} & \makecell{LVMs \\ (meV)} &  &  & & \makecell{$E_\text{ZPL}$ \\ (eV)} & \multicolumn{5}{c}{\makecell{LVMs \\ (meV)}}  \\ \hline
 & & & & & & $(\text{C}_{2})_{\text{Si},h}$ & 1.60 & 114.5 &   &    &    &  \\
 & & & & & & $(\text{C}_{2})_{\text{Si},k}$ & 1.54 & 114.4 &   &    &    &  \\  \hline
 & & & & & & $(\text{C}_\text{BC})_{3,kkk.\text{C}}$ & 2.83  & 117.9 &  124.5 &   145.1 &  154.7  &  \\  \hline
\multirow{3}*{\makecell{triplet \\ emitter \\ in 4H-SiC~\citep{steeds2008creation,steeds2008identification}}}  & 463.3 & 2.68 & 132.82 & 179.86 & &  $(\text{C}_\text{sp})_{2,hk.\text{cub}}$  & 2.61 & 129.8 & 178.1 & & & \\ 
    & 463.6 & 2.67 & 132.53 & 178.47 & &  $(\text{C}_\text{BC})_{2,hh}$  & 2.56 & 129.9 & 177.2 & & &  \\
    & 464.3 & 2.67 & 131.90 & 180.03 & &  $(\text{C}_\text{BC})_{2,kk}$  & 2.59 & 129.6 & 179.6 & & &  \\ \hline
\multirow{10}*{\makecell{other \\ optical \\ centers \\ in 4H-SiC~\citep{steeds2008identification}}} & 414.2 & 2.99 & 119.9 & 122.5 & 129.2 & $(\text{C}_{sp})_{4,hhkk}$  & 2.89 & 120.2 & 121.4 & 125.8 & 134.7 &   \\ 
 & 450.6 & 2.75 & 135.1 & 180.0 &       &                                            &      &       &   &    &    & \\
 & 456.7 & 2.71 & 133.5 &       &       & $(\text{C}_\text{BC})_{2,hh.\text{plane}}$ & 2.70 & 132.7 &   &    &    &  \\
 & 465.7 & 2.66 & 177.8 &       &       &                                            &      &       &   &    &    & \\
 & 471.8 & 2.63 & 151.8 & 189.4 & 246.9 & \makecell{$(\text{C}_{3})_{\text{Si},h}$ \\ $(\text{C}_{3})_{\text{Si},k}$}  & \makecell{2.67 \\2.56} & \makecell{150.7 \\153.2} &  \makecell{246.7 \\248.6} &    &    & \\
 & 493.5 & 2.51 & 166.9 & 169.8 &       &                                    &      &       &       &       &       & \\
 & 498.5 & 2.49 & 179.9 &       &       &                                    &      &       &       &       &       & \\
 & 520.6 & 2.38 & 122.1 & 172.3 &       &                                    &      &       &       &       &       & \\
 & 596.8 & 2.08 & 123.9 & 149.1 &       &  $(\text{C}_\text{BC})_{4,hhkk}$   & 2.20 & 121.7 & 127.4 & 131.7 & 146.3 & 149.3 \\
 & 599.3 & 2.07 & 124.4 & 160.5 & 180.5 &                                    &      &       &       &       &       & \\ \hline
\multirow{4}{2cm}{optical centers in $\text{4H-SiC}/\text{SiO}_2$~\citep{johnson2019optically}} & 584.2 & 2.12 & \multicolumn{2}{c}{\makecell{150$\sim$200}}   &      &      &       &   &    &    & \\ 
  & 585.0 & 2.12 & \multicolumn{2}{c}{\makecell{120$\sim$200}}   &      & $(\text{C}_\text{BC})_{4,kkkk}$  & 2.36 & 120.5 & 122.1 & 126.0 & 145.5 & 192.8  \\
  & 601.5 & 2.06 & \multicolumn{2}{c}{\makecell{180$\sim$200}}   &      &      &       &   &    &    & \\
  & 615.0 & 2.02 & \multicolumn{2}{c}{\makecell{120$\sim$200}}   &      &      &       &   &    &    & \\ 
\end{tabular}
\end{ruledtabular}
\end{table*}
%%%% table 6 %%%%

The \textit{U} center was previously associated with the triscele structure of the tri-carbon antisite clusters [Fig.~\ref{Figure 7}(a)] which can only produce a very high LVM at around 247~meV~\cite{steeds2008identification, mattausch2006thermally}. Our calculations agree with this conclusion as these defects in 4H-SiC indeed produce LVMs at around 150~meV and 247~meV in the PL spectrum (see Table~\ref{Table 6}). The calculated ZPL energy at around 2.6~eV agrees well with the 471.8-nm emitter in 4H-SiC. However, an additional LVM at around 190~meV was associated with this PL center~\cite{steeds2008identification}. We think that that feature in the PL spectrum was falsely associated with this PL center and it is likely a part of another PL center with overlapping spectra of the 471.8-nm emitter. The $U$ center in 6H-SiC has $E_\text{ZPL}=2.36$~eV which follows well the CBM shifts between 4H-SiC and 6H-SiC with reference to the 471.8-nm emitter at $E_\text{ZPL}=2.63$~eV. This further supports our proposal that the two defects are common and should have similar LVM structure in the PL spectrum. 

Along with the \textit{U} center, the \textit{Z} center is another optical center in 6H-SiC that shows an increase in concentration during annealing~\cite{mattausch2006thermally}. The \textit{Z} center shows considerable formation following annealing at 900~$^{\circ}$C and remains stable up to 1300~$^{\circ}$C. The 465.7-nm emitter in 4H-SiC displays a similar ZPL to that of the \textit{Z} center considering the CBM difference between 4H-SiC and 6H-SiC, whereas its highest LVM is lower than that of the latter. Given that the formation of the \textit{Z} center with annealing temperature is similar to that of the \textit{U} center, we speculate that the \textit{Z} center may also originate from a tri-carbon or larger carbon clusters.

Furthermore, the analysis of ZPLs and LVMs suggests that the optical properties of tetra-carbon interstitials $(\text{C}_\text{sp})_{4,hhkk}$ and $(\text{C}_\text{BC})_{4,hhkk}$ could be associated to those of 414.2-nm and 596.8-nm emitters in 4H-SiC, respectively (Table~\ref{Table 6}). Annealing experiments in Ref.~\onlinecite{steeds2008identification} revealed that the concentration of the 596.8-nm emitter is higher than that of the 414.2-nm emitter at annealing temperatures below 1100~$^{\circ}$C, whereas at higher temperatures, the 414.2-nm emitter is favored. This behavior can be explained by the formation energies of $(\text{C}_\text{sp})_{4,hhkk}$ and $(\text{C}_\text{BC})_{4,hhkk}$ (c.f., Table~\ref{Table 3}). Another stable form of tetra-carbon interstitial defect [$(\text{C}_\text{BC})_{4,kkkk}$] generates a distinguishable PL spectrum (Table~\ref{Table 6}), sharing a similar optical characteristic with the optical emitters observed at the $\text{4H-SiC}/\text{SiO}_2$ interface, which consist of sharp high-energy LVMs between 120 and 200~meV~\cite{johnson2019optically}. Although, the calculated ZPL of $(\text{C}_\text{BC})_{4,kkkk}$ is 2.36~eV, slightly higher than that of the optical emitters at the $\text{4H-SiC}/\text{SiO}_2$ interface, but this could be attributed to the amorphous local geometry around the emitters. During the growth or annealing of the $\text{SiO}_2$ layer on SiC, the high carbon solubility of carbon atoms in SiC can result in segregation at the interface, leading to the formation of larger carbon clusters. Thus, the optical emitters at the $\text{4H-SiC}/\text{SiO}_2$ interface may be associated with these larger carbon clusters.       
   
We predict that all the reported PL centers possess bound exciton excited states according to our predictions that may be confirmed in future PLE studies with tunable lasers in the appropriate wavelength region.

We find that vast majority of the carbon-clusters produce deep filled levels close to the valence band maximum which produces fluorescence at significantly higher ZPL energies than the ZPL energies of the divacancies (around 1.1~eV) and Si-vacancies (around 1.4~eV) qubits. On the other hand, the (C$_2$)$_\text{Si}$, di-carbon antisite defects can be presumably excited by 785~nm laser that is often used for off-resonant excitation of divacancy and Si-vacancy qubits. These defects produce multiple donor and an acceptor levels in the band gap [see Fig.~\ref{Figure 2}(c)], thus they can also act as a sink and source of carriers upon illumination and may contribute to the charge fluctuation around the target defect qubits. The relatively low formation energies of these defects imply that they are thermally stable. We provide the fluorescence spectrum of the neutral defects [Fig.~\ref{Figure 7}(b)] that could be observable in the red and near-infrared wavelength regions.

The di-carbon antisite defects may be identified by electron spin resonance techniques too as the positively and negatively charged defects have spin-doublet ground state whereas the neutral state has a spin-triplet metastable state that may be thermally occupied at elevated temperatures. The characteristic hyperfine constants are listed in Table~\ref{Table 7} as obtained by HSE06 calculations. The spin densities are mostly localized on the carbon dangling bond but it could be difficult to observe in natural 4H-SiC as the $^{13}$C isotope has a low 1.1\% abundance. The spin density is also quite observable for the first neighbor 4 carbon atoms ($\sim15$\dots$\sim60$~MHz depending on the charge state and the actual ionic position). The hyperfine constants are also well observable for the second neighbor $^{29}$Si isotopes. Among the 12 neighbor atoms, 4 or 9 $^{29}$Si isotopes have much larger absolute hyperfine constants than $10$~MHz depending on the charge state. Since the natural abundance of $^{29}$Si is about 4.5\% the hyperfine sideband in the electron spin resonance spectra from these defects should have a high intensity. The hyperfine sideband pattern from $^{29}$Si isotopes can be used to distinguish the two configurations in 4H-SiC as the second neighbor atoms are more sensitive to the crystal field environment in the $h$ and $k$ sites. For instance, a relatively strong hyperfine signal is detected for a fifth $^{29}$Si isotope in the positive charge state of (C$_2$)$_{\text{Si}, h}$ which is missing for (C$_2$)$_{\text{Si}, k}$. 

The results of $(\text{C}_{2})_{\text{Si},h}$ and $(\text{C}_{2})_{\text{Si},k}$ in the negative charge state are consistent with the experimental results of HEI6 and HEI5 EPR centers~\cite{umeda2009dicarbon}, respectively, which reaffirms the previous identification of HEI5 and HEI6 EPR centers with the dicarbon antisite defects in the negative charge state~\cite{umeda2009dicarbon} (Table~\ref{Table 7}). We note that the HEI5 and HEI6 EPR centers are annealed out at 1000~$^{\circ}$C which shows a high thermal stability of these defects~\cite{umeda2009dicarbon}.

The zero-field-splitting parameters of the spin triplet are also characteristic for the configurations in the neutral charge state. We obtain $D=-2111$~MHz, $E=14$~MHz and $D=-2174$~MHz, $E=63$~MHz for $h$ and $k$ configurations, respectively, with assuming dominant electron spin-electron spin dipole-dipole interaction. Since a shelving singlet state lies between the triplet states intersystem crossing may occur resulting in spin-selective fluorescence and optical spinpolarization of these defects that could turn these defects to be employed as qubits when can be isolated in 4H-SiC. Exploring the fine electronic structure of these defects is an interesting avenue along this direction but it is out of the scope of the present study.

%%%% table 7 %%%%
\begin{table*}[htb]
\caption{\label{Table 7} Calculated hyperfine constants in MHz unit for the (C$_2$)$_\text{Si}$ defects in their positive, neutral and negative charge states as obtained from HSE06 calculations. The core polarization contribution to the Fermi-contact term is included. $\text{C}_\text{d1}$ and $\text{C}_\text{d2}$ are the defect carbon atoms, $\text{C}_\text{n1}$-$\text{C}_\text{n4}$ are the first nearest neighbors and $\text{Si}_\text{n1}$-$\text{Si}_\text{n12}$ are the second nearest neighbors. The experimental results of HEI5 and HEI6 EPR centers are also listed for comparison~\cite{umeda2009dicarbon}.}
\begin{ruledtabular}
\begin{tabular}{c|rrr|rrr|rrr|r}
 & \multicolumn{3}{c}{\makecell{$(\text{C}_{2})_{\text{Si},h}$ positive}} & \multicolumn{3}{c}{\makecell{$(\text{C}_{2})_{\text{Si},h}$ neutral}} & \multicolumn{3}{c}{\makecell{$(\text{C}_{2})_{\text{Si},h}$ negative}} & HEI6~\cite{umeda2009dicarbon} \\\hline
 & $A_{xx}$ & $A_{yy}$ & $A_{zz}$ & $A_{xx}$ & $A_{yy}$ & $A_{zz}$ & $A_{xx}$ & $A_{yy}$ & $A_{zz}$  & \\\hline 
  $\text{C}_\text{d1}$    & -23.46 & -22.43 & -27.35 &   2.19 &   1.16 &  96.32 &  30.78 &  29.77 & 219.03 & 207.09 \\
  $\text{C}_\text{d2}$    &  27.63 &  26.28 & 222.65 &   3.09 &   2.13 &  96.40 & -19.68 & -19.42 & -23.15 &  \\\hline
  $\text{C}_\text{n1}$    &  33.32 &  33.04 &  41.72 &   6.20 &   4.78 &   9.59 & -25.07 & -23.78 & -26.48 &  \\
  $\text{C}_\text{n2}$    &  50.36 &  49.91 &  60.02 &  15.83 &  11.98 &  17.05 & -24.12 & -16.95 & -25.92 &  \\
  $\text{C}_\text{n3}$    & -23.28 & -13.48 & -24.94 &  14.84 &  10.84 &  15.64 &  49.22 &  48.59 &  59.86 &  \\
  $\text{C}_\text{n4}$    & -23.28 &  13.49 & -24.94 &  14.89 &  10.89 &  15.69 &  49.23 &  48.60 &  59.86 & \\\hline
  $\text{Si}_\text{n1}$   &   1.77 &   1.03 &   1.97 & -39.12 & -38.69 & -47.93 & -82.21 & -81.75 & -98.60 & 96.17 \\
  $\text{Si}_\text{n2}$   & -20.42 & -20.38 & -24.52 & -10.86 & -10.85 & -13.14 &  -1.02 &  -0.82 &  -2.24 & \\
  $\text{Si}_\text{n3}$   &   1.77 &   1.04 &   1.98 & -39.09 & -38.65 & -47.87 & -82.21 & -81.75 & -98.59 & 96.17 \\
  $\text{Si}_\text{n4}$   &  -1.52 &  -1.15 &  -4.23 & -33.72 & -32.79 & -40.06 & -71.23 & -69.60 & -82.47 & 77.59 \\
  $\text{Si}_\text{n5}$   &  -1.59 &  -1.21 &  -4.32 & -33.81 & -32.87 & -40.15 & -71.25 & -69.62 & -82.49 & 77.59 \\
  $\text{Si}_\text{n6}$   &  -2.84 &  -2.53 &  -3.70 &  -1.57 &  -1.19 &  -1.68 &  -0.21 &  -0.17 &   0.69 & \\
  $\text{Si}_\text{n7}$   & -52.55 & -51.47 & -65.68 & -28.52 & -27.96 & -35.64 &  -3.76 &  -3.41 &  -5.21 &  \\
  $\text{Si}_\text{n8}$   &  -0.41 &   0.17 &  -0.69 &   0.16 &  -0.02 &   0.44 &   1.87 &   0.70 &   2.15 &  \\
  $\text{Si}_\text{n9}$   & -50.01 & -49.01 & -62.52 & -25.99 & -25.25 & -32.49 &   2.00 &  -0.18 &   2.43 &  \\
  $\text{Si}_\text{n10}$  &  -0.42 &   0.16 &  -0.70 &   0.17 &  -0.01 &   0.45 &   1.88 &   0.71 &   2.15 &  \\
  $\text{Si}_\text{n11}$  & -52.62 & -51.54 & -65.78 & -28.50 & -27.94 & -35.62 &  -3.76 &  -3.41 &  -5.21 &  \\
  $\text{Si}_\text{n12}$  & -49.97 & -48.97 & -62.45 & -25.93 & -25.19 & -32.42 &   2.00 &  -0.18 &   2.43 &  \\  \hline 
   \\ %\hline
   & \multicolumn{3}{c}{\makecell{$(\text{C}_{2})_{\text{Si},k}$ positive}} & \multicolumn{3}{c}{\makecell{$(\text{C}_{2})_{\text{Si},k}$ neutral}} & \multicolumn{3}{c}{\makecell{$(\text{C}_{2})_{\text{Si},k}$ negative}} & HEI5~\cite{umeda2009dicarbon} \\\hline
 & $A_{xx}$ & $A_{yy}$ & $A_{zz}$ & $A_{xx}$ & $A_{yy}$ & $A_{zz}$ & $A_{xx}$ & $A_{yy}$ & $A_{zz}$ & \\\hline 
  $\text{C}_\text{d1}$ 	&  -20.46 	&	-19.34 	&	-24.16 		&	3.67 	&	2.62 	&	96.77 		&	30.51 	&	29.47 	&	215.81 & 210.93 \\
  $\text{C}_\text{d2}$ 	&	45.97 	&	44.44 	&	241.95 		&	9.92 	&	8.94 	&	104.28 		&	-18.88 	&	-18.74 	&	-22.47 &  \\ \hline
  $\text{C}_\text{n1}$	&	51.93 	&	51.18 	&	63.34 		&	17.20 	&	13.27 	&	19.01 		&	-24.24 	&	-16.76 	&	-26.10 &  \\
  $\text{C}_\text{n2}$	&	31.77 	&	31.68 	&	39.01 		&	9.15 	&	4.83 	&	9.80 		&	-21.37 	&	-15.61 	&	-23.63 &  \\
  $\text{C}_\text{n3}$	&	-18.07 	&	-8.70 	&	-19.70 		&	13.10 	&	10.00 	&	16.12 		&	39.81 	&	39.14 	&	 51.80 &  \\
  $\text{C}_\text{n4}$	&	-18.08 	&	-8.71 	&	-19.70 		&	13.05 	&	9.95 	&	16.06 		&	39.82 	&	39.14 	&	 51.80 &  \\ \hline
  $\text{Si}_\text{n1}$ 	&	0.26 	&	-0.28 	&	-1.36 		&	-0.12 	&	-0.08 	&	0.44 		&	-0.53 	&	-0.29 	&  	  0.63 &  \\
  $\text{Si}_\text{n2}$ 	&	-0.96 	&	-0.62 	&	-4.01 		&	-24.29 	&	-23.39 	&	-31.12 		&	-52.44 	&	-51.19 	&	-64.97 & 59.00 \\
  $\text{Si}_\text{n3}$ 	&	-0.96 	&	-0.62 	&	-4.01 		&	-24.27 	&	-23.37 	&	-31.10 		&	-52.44 	&	-51.19 	&	-64.96 & 59.00  \\
  $\text{Si}_\text{n4}$ 	&	-1.55 	&	-1.22 	&	-2.97 		&	-31.99 	&	-31.50 	&	-38.56 		&	-59.37 	&	-58.40 	&	-70.47 & 66.08  \\
  $\text{Si}_\text{n5}$ 	&	-1.55 	&	-1.22 	&	-2.97 		&	-32.03 	&	-31.54 	&	-38.62 		&	-59.38 	&	-58.40 	&	-70.47 & 66.08  \\
  $\text{Si}_\text{n6}$ 	&	-0.90 	&	-0.76 	&	-1.72 		&	-11.38 	&	-10.65 	&	-10.53 		&	1.19 	&	0.45 	&	  1.62 &  \\
  $\text{Si}_\text{n7}$ 	&	-0.75 	&	0.06 	&	-0.99 		&	-0.51 	&	-0.18 	&	-0.63 		&	0.81 	&	-0.32 	&	  1.01 &  \\
  $\text{Si}_\text{n8}$ 	&	-51.39 	&	-50.17 	&	-60.84 		&	-27.83 	&	-26.93 	&	-32.82 		&	-1.62 	&	-1.18 	&	 -3.57 &  \\
  $\text{Si}_\text{n9}$ 	&	-50.65 	&	-49.89 	&	-65.44 		&	-26.84 	&	-26.35 	&	-34.43 		&	0.74 	&	-0.53 	&	  0.99 &  \\
  $\text{Si}_\text{n10}$ 	&	-51.39 	&	-50.17 	&	-60.84 		&	-27.86 	&	-26.97 	&	-32.84 		&	-1.62 	&	-1.19 	&	 -3.57 &  \\
  $\text{Si}_\text{n11}$ 	&	-0.75 	&	0.06 	&	-0.99 		&	-0.51 	&	-0.17 	&	-0.62 		&	0.81 	&	-0.33 	& 	  1.01 &  \\
  $\text{Si}_\text{n12}$ 	&	-50.65 	&	-49.89 	&	-65.44 		&	-26.86 	&	-26.36 	&	-34.45 		&	0.74 	&	-0.53 	&	  0.99 &  \\  
\end{tabular}
\end{ruledtabular}
\end{table*}
%%%% table 7 %%%%

%----------CONCLUSION----------%
\section{Summary and conclusion}
\label{sec:conclusion}
In conclusion, the thermodynamic stability, electronic structure, vibrational properties, and photoluminescence spectrum of carbon-clusters were studied in 4H-SiC by means of \textit{ab initio} density functional theory calculations. The single carbon interstitial and antisite defects prefer to trap carbon atoms and form stable clusters consisting of multiple carbon atoms. Tetra-carbon clusters may only form as a combination of carbon self-interstitials. We identified the $(\text{C}_\text{sp})_{2,hk.\text{cub}}$, $(\text{C}_\text{BC})_{2,hh}$ and $(\text{C}_\text{BC})_{2,kk}$ defects as the origin of the 463-nm triplet emitters, and $(\text{C}_\text{BC})_{2,hh.\text{plane}}$ can be well associated with the 456.6-nm emitter because the calculated and observed ZPL energies and LVMs agree well. $(\text{C}_3)_\text{Si}$ defects are assigned to the 471.8-nm emitter in 4H-SiC. We tentatively associate the tri-carbon interstitial clusters to some observed carbon-related emitters in 4H-SiC and some tetra-interstitial clusters show the features of fluorescence centers detected at 4H-SiC/SiO$_2$ interface. The dicarbon antisite defects are identified as the most harmful to the vacancy-type defect qubits because of their overlapping absorption and emission spectra and electrical activities that may lead to charge switching around the target vacancy-type defect qubits. We provide the characteristic hyperfine constants and zero-field-splitting parameters of the dicarbon antisite defects as reference for future electron spin resonance observations. Our results indicate that the thermally stable dicarbon antisite defects shows potential to turn them from harmful defects to qubit defects based on their magneto-optical properties that might be validated in future studies.

%====================================================
% others %
\section*{Author contribution}
All authors contributed to the discussion and writing the manuscript. AG led the entire scientific project.

\section*{Competing interests}
The authors declare that there are no competing interests.

\section*{Data Availability}
The data that support the findings of this study are available from the corresponding author upon reasonable request.

%----------ACKNOWLEDGEMENT----------%
\begin{acknowledgments}
Support by the National Excellence Program for the project of Quantum-coherent materials (NKFIH Grant No.\ KKP129866) as well as by the Ministry of Culture and Innovation and the National Research, Development and Innovation Office within the Quantum Information National Laboratory of Hungary (Grant No.\ 2022-2.1.1-NL-2022-00004) is much appreciated. AG acknowledges the high-performance computational resources provided by KIF\''U (Governmental Agency for IT Development) institute of Hungary and the European Commission for the project QuanTelCo (Grant No.\ 862721). BH acknowledges the NSFC (Grants Nos.\ 12088101 and 12174404), National Key Research and Development of China (Grant No.\ 2022YFA1402400), NSAF (Grant No.\ U2230402). PL acknowledges the China Postdoctoral Science Foundation (Grant No.\ 2020M680319) and the International Postdoctoral Exchange Fellowship Program (Grant No.\ PC2021009). 
\end{acknowledgments}

%----------APPENDIX----------%
\begin{appendix}  
\setcounter{section}{0}
\setcounter{table}{0}   
\setcounter{figure}{0}
\setcounter{equation}{0}
\renewcommand{\thesection}{\Alph{section}}
\renewcommand{\thetable}{A\arabic{table}}
\renewcommand{\thefigure}{A\arabic{figure}}
\renewcommand{\theequation}{A\arabic{equation}}

\section{Theory and computational methodology of the simulated fluorescence spectrum}
\label{sec:appendix A}
In our calculation, the fluorescence spectrum of the defect is computed within the HR theory~\citep{huang1950theory}. This requires determining the adiabatic potential energy surface of the electronic excited state as well as the phonons that may participate in the optical transitions. The excited state of the defect is calculated by $\Delta$SCF method with hybrid density functional theory~\citep{PhysRevLett.103.186404}. The phonon sideband of the fluorescence spectrum is given by~\citep{stoneham2001theory} 
\begin{equation}
\label{equationA1}
\begin{split}
I(\hbar\omega)= & \frac{n_D\omega^3}{3\varepsilon_0{\pi}c^3\hbar}\vert\vec{\mu}_{eg}\vert^2\times \\ 
                & \sum_m\vert\langle\chi_{gm}\vert\chi_{e0}\rangle\vert^2\delta(E_\text{ZPL}-E_{gm}-\hbar\omega)\text{,}
\end{split}
\end{equation}
\noindent where $n_D$ is the refractive index of the material, $\vec{\mu}_{eg}$ is the electronic transition dipole moment, $\chi_{gm}$ and $\chi_{e0}$ are the vibrational levels of the ground and excited state, $E_\text{ZPL}$ is the energy difference between excited and ground state, $E_{gm}$ is energy of ground state. In order to directly compare with the experimental results, we considered the normalized luminescence intensity which is defined as
\begin{equation}
\label{equationA2}
\begin{split}
L(\hbar\omega)=CA(\hbar\omega){,}
\end{split}
\end{equation}
\begin{equation}
\label{equationA3}
\begin{split}
A(\hbar\omega)=\sum_m\vert\langle\chi_{gm}\vert\chi_{e0}\rangle\vert^2\delta(E_\text{ZPL}-E_{gm}-\hbar\omega)\text{.}
\end{split}
\end{equation}
\noindent $C$ [$C^{-1}=\int{A}(\hbar\omega)\omega^3d(\hbar\omega)$] is the normalization constant, $A(\hbar\omega)$ is the optical spectral function and given as the Fourier transform of the generating function $G(t)$~\citep{alkauskas2014first}
\begin{equation}
\label{equationA4}
\begin{split}
A(E_\text{ZPL}-\hbar\omega)=\frac{1}{2\pi}\int_{-\infty}^\infty{G(t)}e^{i\omega{t}-\gamma\vert{t}\vert}dt\text{.}
\end{split}
\end{equation}
\noindent The generating function $G(t)=e^{(S(t)-S(0))}$. $S(t)$ and $S(0)$ are defined as
\begin{equation}
\label{equationA5}
\begin{split}
S(t)=\int_{0}^\infty{S(\hbar\omega)}e^{-i\omega{t}}d(\hbar\omega)
\end{split}
\end{equation}
\noindent and
\begin{equation}
\label{equationA6}
\begin{split}
S(0)=\int_{0}^\infty{S(\hbar\omega)}d(\hbar\omega)\text{.}
\end{split}
\end{equation}
\noindent In Eqs.(\ref{equationA5}) and (\ref{equationA6}), $S(\hbar\omega)$ is given as
\begin{equation}
\label{equationA7}
\begin{split}
S(\hbar\omega)=\sum_kS_k\delta(\hbar\omega-\hbar\omega_k)\text{,}
\end{split}
\end{equation}
\noindent where $S_k$ is the partial HR factor and given as $S_k=\frac{1}{2\hbar}\omega_kq_k^2$, in which $q_k$ is the mass-weighted difference of the ground state and the excited state geometries, evaluated as
\begin{equation}
\label{equationA8}
\begin{split}
q_k=\frac{1}{\omega_k^2}\sum_{\alpha{i}}\frac{1}{m_\alpha^{1/2}}(F_{e;\alpha{i}}-F_{g;\alpha{i}})\Delta{r_{k;\alpha{i}}}\text{.}
\end{split}
\end{equation}
\noindent $F_{e;\alpha{i}}-F_{g;\alpha{i}}$ is the change of the force on the atom $\alpha$ along the direction $i$ for a fixed position of all atoms. $q_k$ determines the phonon sideband in the photoluminescence spectrum. The total HR factor for a given optical transition is defined as $S=\sum_kS_k$.

\section{Partial Huang-Rhys factors of $(\text{C}_2)_{\text{Si},h}$, $(\text{C}_2)_{\text{Si},k}$, $(\text{C}_3)_{\text{Si},h}$ and $(\text{C}_3)_{\text{Si},k}$}
\label{sec:appendix B}
\begin{figure}[H]
\includegraphics[width=\columnwidth]{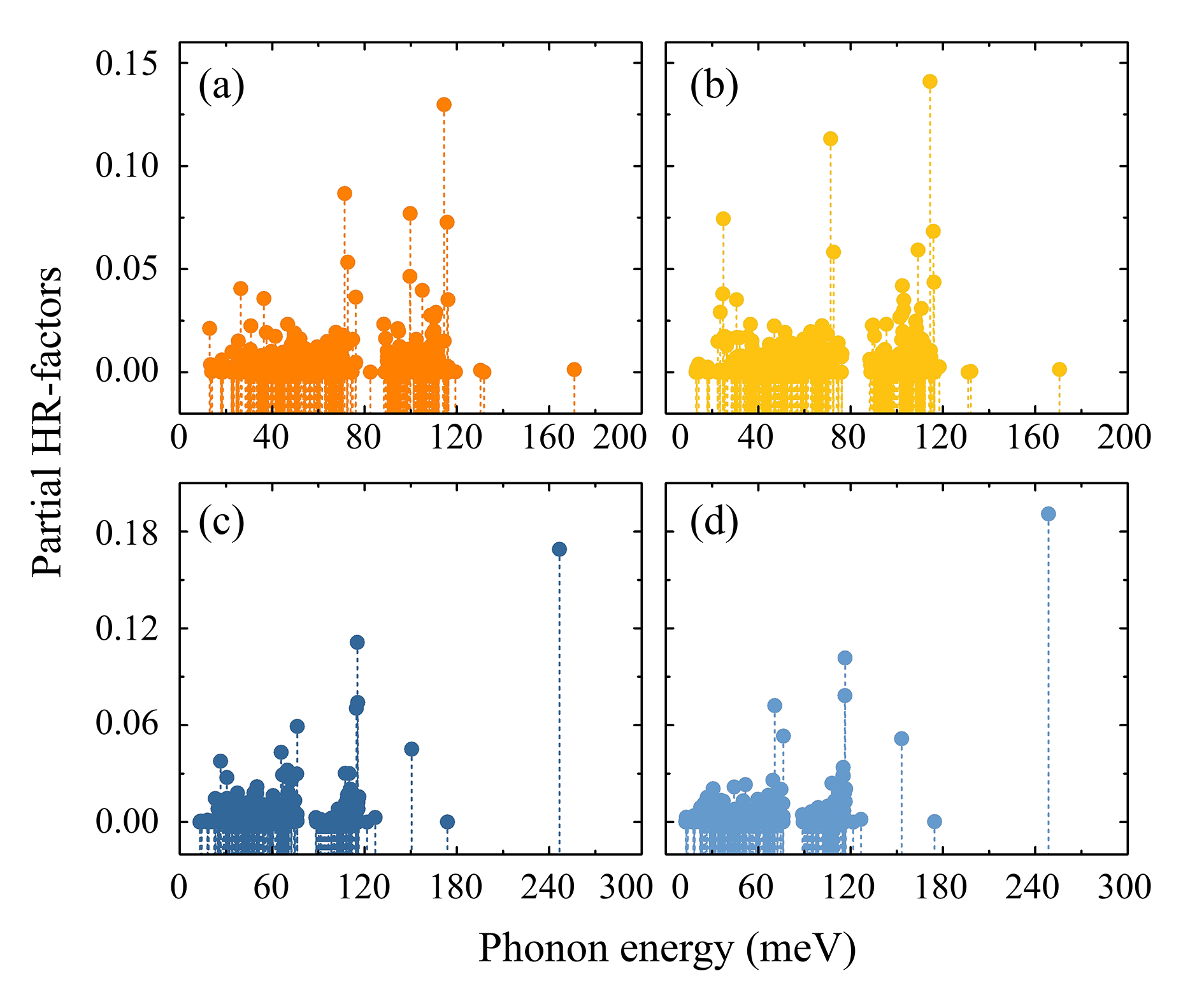}
\caption{\label{Figure A1}%
The partial Huang-Rhys factors of (a) $(\text{C}_2)_{\text{Si},h}$, (b) $(\text{C}_2)_{\text{Si},k}$, (c) $(\text{C}_3)_{\text{Si},h}$ and (d) $(\text{C}_3)_{\text{Si},k}$.}
\end{figure}

The partial Huang-Rhys factors of $(\text{C}_2)_{\text{Si},h}$, $(\text{C}_2)_{\text{Si},k}$, $(\text{C}_3)_{\text{Si},h}$ and $(\text{C}_3)_{\text{Si},k}$ are shown in Fig.~\ref{Figure A1}.  
%%%% figure A1 %%%%
%%%% figure A1 %%%%

\end{appendix}

\bibliography{mainref}

\end{document}